\documentclass[]{article}

\usepackage{arxiv}

\usepackage[utf8]{inputenc} 
\usepackage[T1]{fontenc}    
\usepackage{hyperref}       
\usepackage{url}            
\usepackage{booktabs}       
\usepackage{amsfonts}       
\usepackage{nicefrac}       
\usepackage{microtype}      
\usepackage{lipsum,amsmath}
\usepackage{algorithm,tikz}
\usepackage[]{algpseudocode}
\usepackage{pgfplots}
\usepackage{subcaption}
\usepackage[]{threeparttable}
\usepackage{multirow,bm}

\newcommand{\Am}{\mathbf{A}}
\newcommand{\Bm}{\mathbf{B}}
\newcommand{\bmm}{\mathbf{b}}
\newcommand{\Dm}{\mathbf{D}}

\newcommand{\Dtr}{\mathcal{D}_\mathrm{tr}}
\newcommand{\Dval}{\mathcal{D}_\mathrm{val}}
\newcommand{\Em}{\mathbf{E}}

\newcommand{\Gm}{\mathbf{G}}
\newcommand{\Hb}{\mathbf{H}_\mathbf{B}}
\newcommand{\Hl}{\mathbf{H}_\mathbf{L}}

\newcommand{\Lm}{\mathbf{L}}
\newcommand{\Lms}{\mathbf{L}_\mathrm{s}}
\newcommand{\Mm}{\mathbf{M}_\mathrm{s}}
\newcommand{\mm}{\mathbf{m}}
\newcommand{\Ntr}{{N}_\mathrm{tr}}
\newcommand{\Nval}{{N}_\mathrm{val}}
\newcommand{\ngm}{n_\mathrm{g}}

\newcommand{\ny}{n_\mathrm{y}}
\newcommand{\nt}{n_\mathrm{t}}
\newcommand{\nw}{n_\mathrm{w}}
\newcommand{\ppm}{\mathbf{p}}
\newcommand{\pnn}{\mathbf{p}_{\!_\mathrm{NN}}}
\newcommand{\Qy}{Q_\mathrm{y}}
\newcommand{\rr}{\mathbf{r}}
\newcommand{\Km}{\mathbf{K}_\mathrm{s}}

\newcommand{\kpost}{k_\mathrm{post}}
\newcommand{\kpre}{k_\mathrm{pre}}
\newcommand{\Cm}{\mathbf{C}}
\newcommand{\Cms}{\mathbf{C}_\mathrm{s}}

\newcommand{\gm}{\mathbf{g}}
\newcommand{\sm}{\mathbf{s}}
\newcommand{\wm}{\mathbf{w}}
\newcommand{\um}{\mathbf{u}}
\newcommand{\vm}{\mathbf{v}}
\newcommand{\Wm}{\mathbf{W}}
\newcommand{\Xm}{\mathbf{X}}
\newcommand{\Xmb}{\overline{\mathbf{X}}}
\newcommand{\xmb}{\overline{\mathbf{x}}}
\newcommand{\Ym}{\mathbf{Y}}
\newcommand{\xm}{\mathbf{x}}
\newcommand{\xo}{\overline{\xm}}
\newcommand{\xnl}{\mathbf{x}_\mathrm{corr}}

\newcommand{\ym}{\mathbf{y}}
\newcommand{\zm}{\mathbf{z}}

\newcommand{\drm}{\mathrm{d}}

\newcommand{\xii}{\boldsymbol{\xi}}
\newcommand{\thetaa}{\boldsymbol{\theta}}
\newcommand{\sigmaa}{\boldsymbol{\sigma}}

\newcommand{\pinn}{\mathcal{M}_\mathrm{NN}}

\newcommand\Tstrut{\rule{0pt}{2.6ex}}         

\usepackage{tikz}
\usetikzlibrary{matrix,chains,positioning,decorations.pathreplacing,arrows}
\usetikzlibrary{positioning,calc}
\usetikzlibrary{patterns,decorations.pathmorphing,decorations.markings}
\usetikzlibrary{shapes}
\usetikzlibrary{fit}
\tikzstyle{block} = [draw,rectangle,thick,minimum height=2em,minimum width=2em]
\tikzstyle{sum} = [draw,circle,inner sep=0mm,minimum size=2mm]
\tikzstyle{connector} = [->,thick]
\tikzstyle{line} = [thick]
\tikzstyle{branch} = [circle,inner sep=0pt,minimum size=1mm,fill=black,draw=black]
\tikzstyle{guide} = []
\tikzset{%
  every neuron/.style={
    circle,
    draw,
    fill = blue!20,
    thick,
    minimum size=1.1cm
  },
  neuron missing/.style={
    draw=none, 
    scale=4,
    fill = none,
    text height=0.333cm,
    execute at begin node=\color{black}$\vdots$
  },
  input neuron/.style={
    circle,
    draw,
    thick,
    fill = yellow!20,
    minimum size=1cm
  },
  output neuron/.style={
    circle,
    draw,
    thick,
    fill = red!20,
    minimum size=1cm
  },
}

\usetikzlibrary{3d,decorations.text,shapes.arrows,positioning,fit,backgrounds}
\tikzset{pics/fake box/.style args={
#1 with dimensions #2 and #3 and #4}{
code={
\draw[gray,ultra thin,fill=#1]  (0,0,0) coordinate(-front-bottom-left) to
++ (0,#3,0) coordinate(-front-top-right) --++
(#2,0,0) coordinate(-front-top-right) --++ (0,-#3,0) 
coordinate(-front-bottom-right) -- cycle;
\draw[gray,ultra thin,fill=#1] (0,#3,0)  --++ 
 (0,0,#4) coordinate(-back-top-left) --++ (#2,0,0) 
 coordinate(-back-top-right) --++ (0,0,-#4)  -- cycle;
\draw[gray,ultra thin,fill=#1!80!black] (#2,0,0) --++ (0,0,#4) coordinate(-back-bottom-right)
--++ (0,#3,0) --++ (0,0,-#4) -- cycle;
\path[gray,decorate,decoration={text effects along path,text={}}] (#2/2,{2+(#3-2)/2},0) -- (#2/2,0,0);
}
}}
\tikzset{circle dotted/.style={dash pattern=on .05mm off 2mm,
                                         line cap=round}}

\title{Uncertainty Quantification of Locally Nonlinear Dynamical Systems using Neural Networks}

\author{
  Subhayan De \\
  Aerospace Engieering Sciences\\
  University of Colorado\\
  Boulder, CO 80309 \\
  \texttt{Subhayan.De@colorado.edu} \\
}

\begin{document}
\maketitle

\begin{abstract}

Models are often given in terms of differential equations to represent physical systems. In the presence of uncertainty, accurate prediction of the behavior of these systems using the models requires understanding the effect of uncertainty in the response. In uncertainty quantification, statistics such as mean and variance of the response of these physical systems are sought. To estimate these statistics sampling-based methods like Monte Carlo often require many evaluations of the models' governing differential equations for multiple realizations of the uncertainty. However, for large complex engineering systems, these methods become computationally burdensome as the solution of the models' governing differential equations for such systems is expensive. In structural engineering, often an otherwise linear structure contains spatially local nonlinearities with uncertainty present in them. A standard nonlinear solver for them with sampling-based methods for uncertainty quantification incurs significant computational cost for estimating the statistics of the response. 
To ease this computational burden of uncertainty quantification of large-scale locally nonlinear dynamical systems, a method is proposed herein, which decomposes the response into two parts --- response of a nominal linear system and a corrective term. This corrective term is the response from a pseudoforce that contains the nonlinearity and uncertainty information. 
In this paper, neural network, a recently popular tool for universal function approximation in the scientific machine learning community due to the advancement of computational capability as well as the availability of open-sourced packages like PyTorch and TensorFlow is used to estimate the pseudoforce. Since only the nonlinear and uncertain pseudoforce is modeled using the neural networks the same network can be used to predict a different response of the system and hence no new network is required to train if the statistic of a different response is sought. 
Three numerical examples are used to show that the proposed method inexpensively produces accurate statistics of the response in the presence of uncertainty. 

\end{abstract}


\section{Introduction}

Models often given by a set of differential equations are used to characterize and express the behavior of a physical system. In these models, the sources of uncertainty can be large, \textit{e.g.}, in material properties, geometry, and loading conditions \cite{hasselman2001quantification,uncertainty2006,bulleit2008uncertainty}. An accurate and robust prediction of the behavior of the physical system using these models requires proper understanding of the effects of these multiple sources of uncertainty. 
Uncertainty quantification using the standard Monte Carlo approach uses many evaluations of the physical system for different realizations of the uncertainty. However, for large and complex structures, this approach soon becomes computationally expensive. Approaches using polynomial chaos expansion \cite{ghanem2003stochastic,xiu2002wiener} and stochastic collocation \cite{babuvska2007stochastic,nobile2008sparse} develop polynomial approximations to reduce the computational burden. 
However, with increasing dimension of the uncertain variables, the number of terms retained in the expansion increases significantly. Similarly, Gaussian process regression \cite{williams2006gaussian} can be used to develop surrogate models \cite{forrester2007multi} but its training cost increases cubically with the number of data points. The response surface approximations \cite{romero2004construction,giunta2006promise} build surrogate models utilizing random samples from the uncertainty but can lead to pitfalls in the presence of a small training sample size \cite{giunta2006promise}. 
Intelligent sampling techniques (\textit{e.g.}, Latin hypercube sampling \cite{mckay2000comparison},  stratified sampling \cite{ross1990course}) can also be implemented for uncertainty quantification of the response of the physical system for a relatively smaller number of realizations of the uncertainty variables. However, the use of surrogate models with these sampling techniques is straightforward and will provide similar reduction in computational cost for all surrogate modeling techniques. 


In structural engineering, an otherwise linear structure often contains spatially local nonlinearities. For example, a building superstructure or a bridge, which behaves linearly under most earthquake or wind excitation, may have a nonlinear base isolation layer \cite{kamalzare2015efficient,de2015fast,de2018computationally} or nonlinear tuned-mass damper attached to it \cite{de2017efficient}. Similarly, spacecrafts often have nonlinear joints \cite{ferri1988modeling,bowden1990joint,krack2017vibration}. Another example of the presence of  local nonlinearity in an otherwise linear structure is contact friction in linear elastic structures \cite{krack2017vibration,ying1992analysis,lee2004dynamic,poudou2007modeling}. For such large-scale locally nonlinear structures, unless an approximate linearization technique is used the computational cost of using a nonlinear solver for uncertainty quantification becomes unbearable \cite{de2017efficient}. \cite{wojtkiewicz2011efficient} used an approach for uncertainty quantification of locally nonlinear dynamical systems that solves a nonlinear Volterra integral equation for multiple realizations of the uncertainty. \cite{de2015efficienta,de2015efficientb,de2017efficient,de2018computationally} used this approach for Bayesian model selection and design under uncertainty of nonlinear structural systems. 

With the availability of highly optimized open-sourced codes like PyTorch \cite{paszke2017automatic} and TensorFlow \cite{abadi2016tensorflow} neural networks have found recent popularity in the scientific community \cite{baker2019workshop}. In these Scientific Machine Learning (SciML) applications, neural networks are used for modeling large complex systems for which response prediction is computationally expensive. In \cite{raissi2017physics,raissi2018hidden,raissi2018deep,raissi2019physics,lu2019deepxde}, the neural networks are trained using a loss function that specifically incorporates the error in the governing differential equations. Using this strategy, the trained networks match the prediction from the governing equations while reducing the computational cost of solving them using methods like finite element.  
\cite{king2018deep} used time-depended recurrent neural networks to learn turbulence. Generative adversarial networks \cite{goodfellow2016nips} are used in \cite{stengel2019physics} to generate high-resolution meteorological data from low-resolution images. Neural networks are also used for reduced order modeling in \cite{hesthaven2018non,wang2019non}. 
\cite{holland2019towards,bhatnagar2019prediction} used neural networks for modeling turbulence by augmenting Reynolds Averaged Navier-Stokes (RANS) models. \cite{wang2019prediction} used neural networks for modeling flow near the wall in Large Eddy Simulation (LES). For estimating structural response, \cite{zhang2019physics,zhang2019deep} used convolutional and recurrent neural networks. Multifidelity datasets for response of physical systems are used for training neural networks in \cite{de2020transfer,meng2020composite,motamed2019multi,chakraborty2020transfer}.
Neural networks have also been used for uncertainty quantification of physical systems. For example,
\cite{zhang2019quantifying} and \cite{gal2016dropout} used dropout strategy \cite{hinton2012improving} for quantifying model as well as parametric uncertainty with neural networks, where some of the connections in the networks are ignored with some probability. \cite{luo2019deep} used convolutional neural networks for developing surrogate models for uncertainty propagation through random field.  \cite{duraisamy2019turbulence} used neural networks to reduce the uncertainty associated with RANS models. 
\cite{de2020transfer} used different transfer learning techniques for uncertainty quantification of physical systems, when training data from an inaccurate coarse model is abundant compared to training data from an accurate fine model. Recently, \cite{chakraborty2020simulation} used this approach of training neural networks for reliability estimation. 

In this paper, to reduce the computational burden associated with a nonlinear solver for a locally nonlinear dynamical system under uncertainty the response is decomposed into two parts --- a nominal linear response and response from a pseudoforce that takes into account the nonlinearity and any uncertainty. The solution for the pseudoforce, however, leads to a nonlinear Voletrra equation of the second kind written in nonstandard form \cite{wojtkiewicz2011efficient}. Hence, to obtain the response for these systems with nonlinearities encountered in structural engineering requires an iterative solver. 
In this paper, instead, a neural network is trained to accurately predict the pseudoforce. The nominal linear response of the dynamical system is then combined with the pseudoforce response to get an accurate estimate of the total response of the system. Hence, this approach is different than the training method of using physics-informed loss function to satisfy the governing equations. Instead, only a part of the response is modeled in this proposed approach keeping the physics solution for rest of the structure. 
Further, the same trained network can be used even if the statistics of a different response is sought as these networks model the pseudoforce and not the response directly. 
Three numerical examples with increasing degrees of freedom (DOF) are used to illustrate the proposed approach. The first example considers uncertainty in a two DOF spring-mass-damper model. The second example uses a 11-story 2-bay 100 DOF building resting on a hysteretic base isolation layer with uncertain properties and subjected to a historic earthquake excitation. A three-dimensional 1623 DOF building with three uncertain tuned mass dampers (TMDs) on its roof is used in the third example. 
These three numerical examples show that once trained the neural networks used in the proposed approach provide accurate prediction along with large computational gains in uncertainty quantification.

\section{Background}
A brief background on uncertainty quantification using surrogate models is discussed in this section followed by a description of the three different architectures for neural networks used in the numerical examples of this paper. The training of these networks using generated datasets is discussed next. 

\subsection{Uncertainty Quantification} 

Dynamical systems are often represented by models given by differential equations. In the presence of uncertainty, the system's response $\Ym(t;\xii)\in\mathbb{R}^{\ny}$ depends on the external force $\wm(t)\in\mathbb{R}^{\nw}$ as well as on the random variables $\xi\in\mathbb{R}^{n_{\xii}}$ and can be given by
\begin{equation}\label{eq:model}
    \Ym(t;\xii) = \mathcal{M}(\wm(t),\xii),
\end{equation}
where $\mathcal{M}:\mathbb{R}^{\nw}\times\mathbb{R}^{n_{\xii}}\rightarrow\mathbb{R}^{\ny}$ is the model of the dynamical system. As a result, $\Ym(t;\xii)$ is also an uncertain quantity. 
In this paper, the random variables $\xii$ are described using known probability distributions. In uncertainty quantification, statistics of the random variables $\Ym(t;\xii)$ are sought. The most commonly used approach for estimating such statistics is the Monte Carlo method \cite{hammersley2013monte}. For example, the mean and variance of $\Ym(t;\xii)$ can be approximated using $N$ realizations of $\xii$ as follows
\begin{equation}\label{eq:mc_eval}
\begin{split}
        \mathbb{E}_{\xii}[\Ym(t;\xii)] &\approx \frac{1}{N}\sum_{i=1}^N \Ym(t;\xii_i);\\
        \mathrm{Var}_{\xii}[Y_j(t;\xii)] &\approx \frac{1}{N-1}\sum_{i=1}^N \left(Y_j(t;\xii_i)-\frac{1}{N}\sum_{k=1}^N Y_j(t;\xii_k)\right)^2;\quad j=1,\dots,\ny,\\
\end{split}
\end{equation}
where the model $\mathcal{M}$ may need to be evaluated for $\{\xii_i\}_{i=1}^N$ for a large $N$ increasing the computational cost. Surrogate models can be developed in such cases \cite{sudret2017surrogate}, which are computationally inexpensive and can be used in \eqref{eq:mc_eval}. In this paper, neural networks are used to replace some parts of the model $\mathcal{M}$ while satisfying the governing equations. 
Note that an intelligent sampling technique can be used instead of \eqref{eq:mc_eval}. However, the focus of this paper is to develop a computationally advantageous strategy for the calculation of $\Ym(t;\xii)$. Same strategy can be applied in conjunction with any other sampling techniques. Hence, the study of different intelligent sampling methods is beyond the scope of this work.

\subsection{Neural Networks}
An artificial neural network, or simply a neural network is widely used for approximating functional relationship such as \eqref{eq:model}. 
With recent advancement in the computing power large neural networks are possible to train that can learn the behavior of complex systems. Among many available architectures for the networks the feed-forward, residual, and convolutional neural networks are used in this paper. They are briefly described next. 

\subsubsection{Feed-forward Neural Network (FNN)}
\begin{figure}[!htb]
\centering
\includegraphics[scale=0.84]{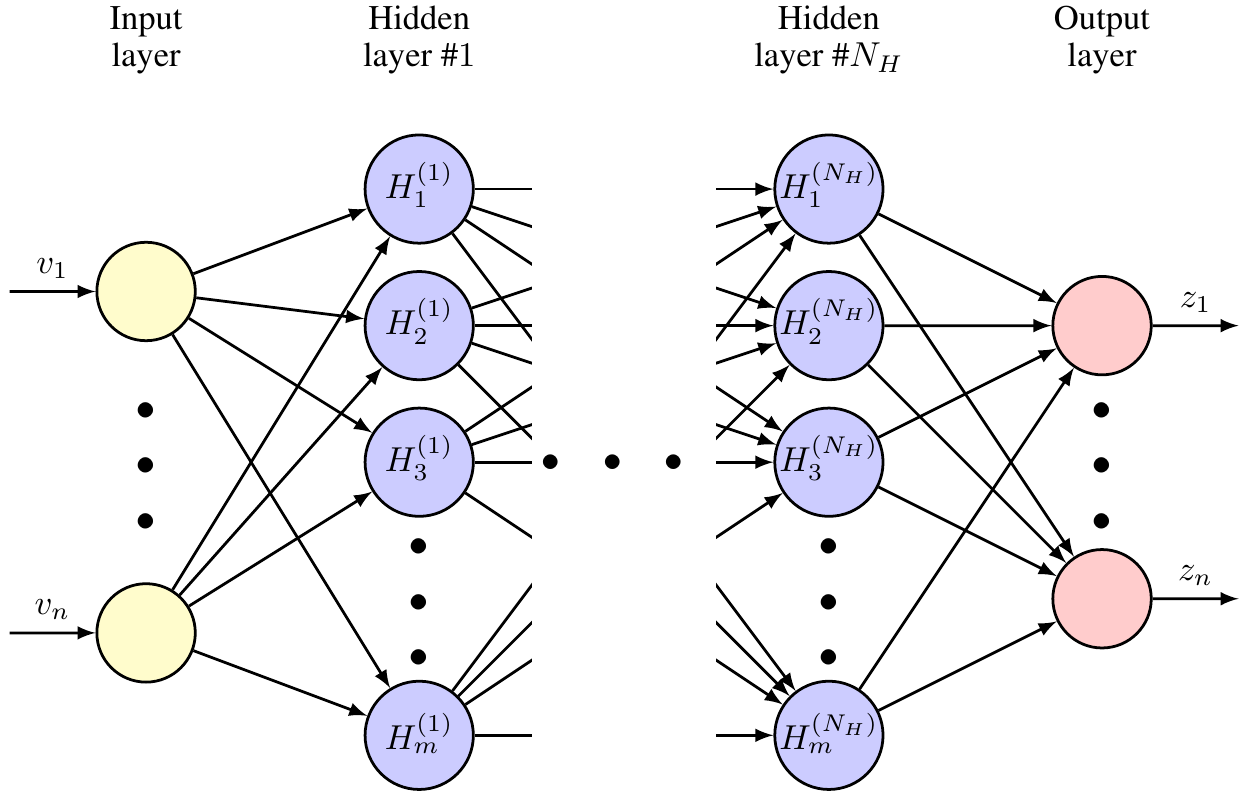}
\caption{Feed-forward Neural Network (FNN) architecture with $N_H$ hidden layers. Each of these hidden layers contains $m$ neurons. }\label{fig:ffn}
\end{figure}

\begin{figure}[!htb]
\centering
\includegraphics[scale=1.0]{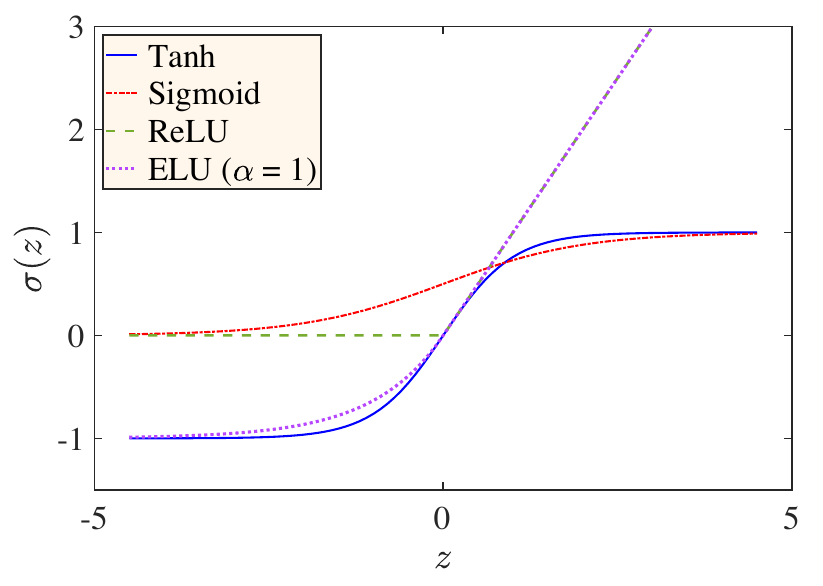}
\caption{Four different activation functions tested in this paper. } \label{fig:activation}
\end{figure}

The feed-forward neural network (FNN) or multilayer perceptron (MLP) \cite{goodfellow2016deep}, which consists of an input layer, one or more hidden layers, and an output layer, is commonly used in scientific machine learning applications \cite{baker2019workshop}. Figure \ref{fig:ffn} shows a schematic of such network, where the network has $N_H$ number of hidden layers. Each of these hidden layers has $m$ number of neurons. However, in general, the number of neurons in each layer can be different. Inside each neuron, an affine transformation followed by a nonlinear activation function is applied to the input. 
Hence, given an input $\vm$ a neural network approximates the output $\zm$ by 
\begin{equation}\label{eq:nn}
\begin{split}
\zm &\approx \pinn\left(\vm;\{\Wm_i\}_{i=0}^{N_H}, \{\bmm_i\}_{i=0}^{N_H} \right)\\
 &= \Wm_0^T\sigmaa_{N_H}(\dots(\sigmaa_1(\Wm_1^T\xm+\bmm_1))\dots)+\bmm_0,\\
\end{split}
\end{equation} 
where $N_H$ is the number of hidden layers; the weights $\Wm_i$ and the biases $\bmm_i$ for $i=1,\dots,N_H$ are the parameters for the $i$th hidden layer that needs to be tuned using the training dataset; $\Wm_0$ and $\bmm_0$ are the parameters for the output layer; and $\sigmaa_i(\cdot)$ is a nonlinear activation function for the $i$th hidden layer. 
There are many choices available for the activation function. For example, the output from a hyperbolic tangent and sigmoid activation functions are, respectively, given by
\begin{equation}\label{eq:act1}
\begin{split}
    \sigma_{\!_{\mathrm{tanh}}}(z)&=\tanh{(z)};\\
    \sigma_{\!_{\mathrm{sigm}}}(z)&=\frac{1}{1+e^{-z}}.\\
\end{split}
\end{equation}
The output from two other popular activation functions, namely, Rectified Linear Unit (ReLU) and Exponential Linear Unit (ELU) are given by
\begin{equation}\label{eq:act2}
\begin{split}
    \sigma_{\!_{\mathrm{ReLU}}}(z)&=\max(0,z);\\
    \sigma_{\!_\mathrm{ELU}}(z) &= \begin{cases}
    z \qquad \qquad \quad~~~\! \text{for } z>0,\\
    \alpha (e^z-1) \qquad \text{for } z\leq0,\\
    \end{cases}\\
\end{split}
\end{equation}
where $\alpha$ is a positive parameter. 
In this paper, the activation functions are chosen from preliminary runs to produce the smallest validation errors defined in Section \ref{sec:datasets}. Figure \ref{fig:activation} compares the outputs of these activation functions. 

\begin{figure}[!htb]
\centering
\includegraphics[scale=0.84]{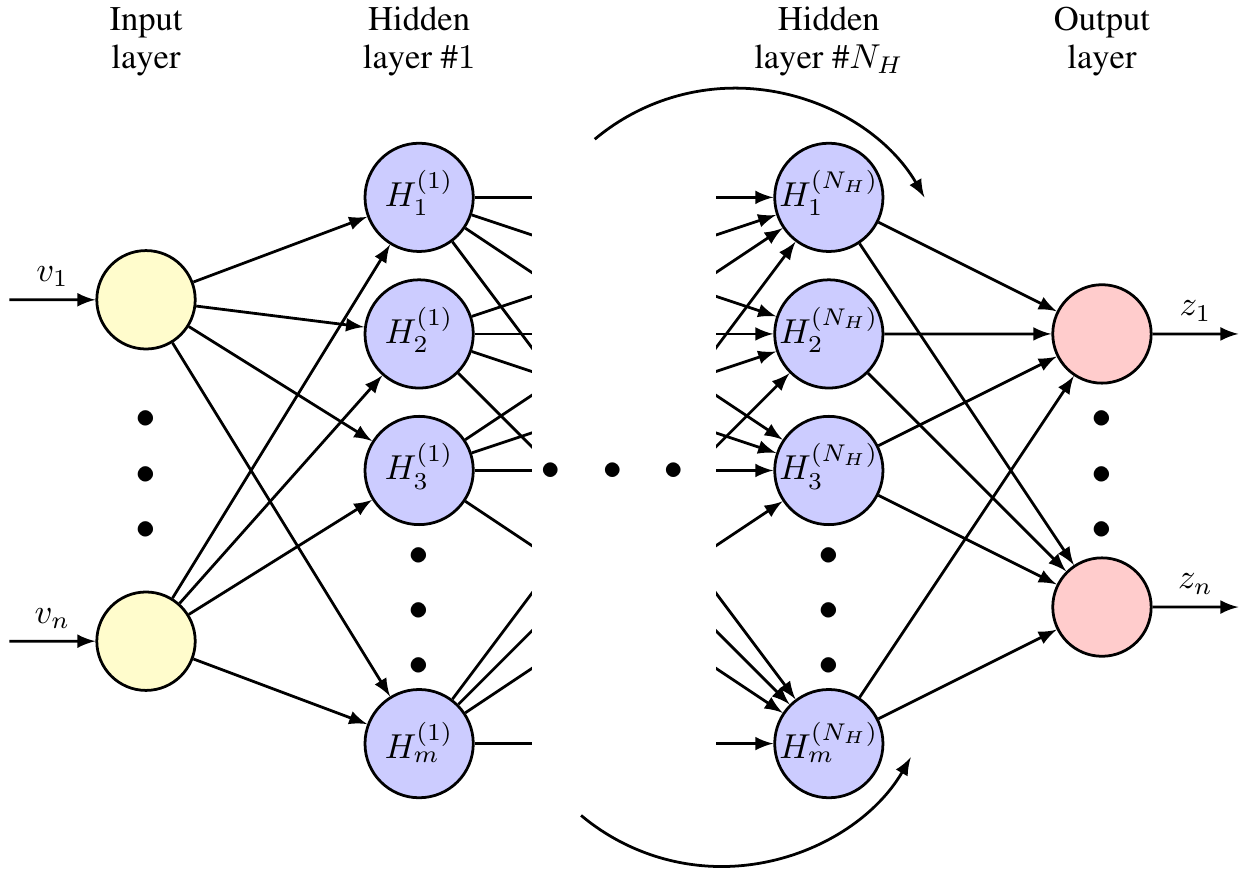}
\caption{A schematic of the Residual Neural Network (ResNet) with residual connections shown using curved arrows. The network here has $N_H$ number of hidden layers, where each of these layers has $m$ neurons.}\label{fig:resnet}
\end{figure}

Another popular neural network architecture is tried herein, where the $i$th hidden layer models the residual error in the output from the previous layer as shown in Figure \ref{fig:resnet} with curved arrows. This architecture is known as the residual neural network or ResNet. Hence, the output of a hidden layer that is used to model the residual after $(i-1)$th layer is given by \cite{he2016deep} 
\begin{equation}\label{eq:resnet}
\zm_i=\sigmaa_i(\Wm_i^T\zm_{i-1}+\bmm_i)+\zm_{i-1}.
\end{equation}
A short-cut mapping \cite{he2016deep} must be used if the dimensions of $\zm_i$ and $\zm_{i-1}$ are different. 

\subsubsection{Convolutional Neural Network (CNN)}\label{sec:cnn}
\begin{figure}
    \centering
    \includegraphics[scale=1.0]{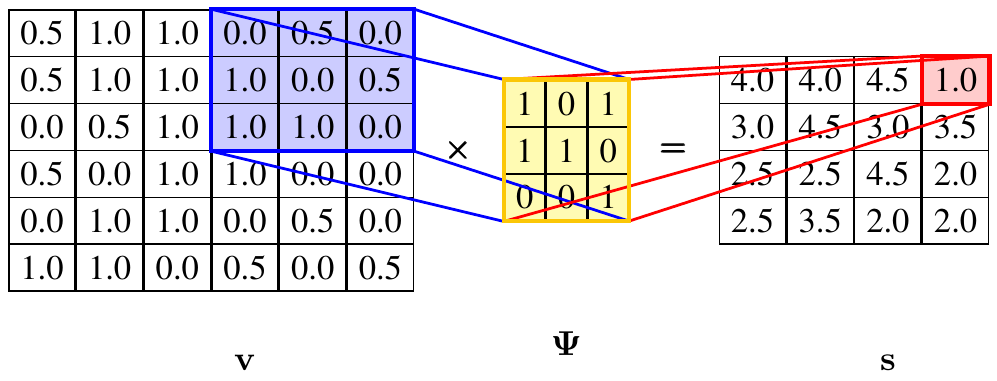}
    \caption{A schematic of the convolution procedure implemented in CNN (see \eqref{eq:crosscorr}).}
    \label{fig:conv}
\end{figure}
\begin{figure}
    \centering
    \includegraphics[scale=1.0]{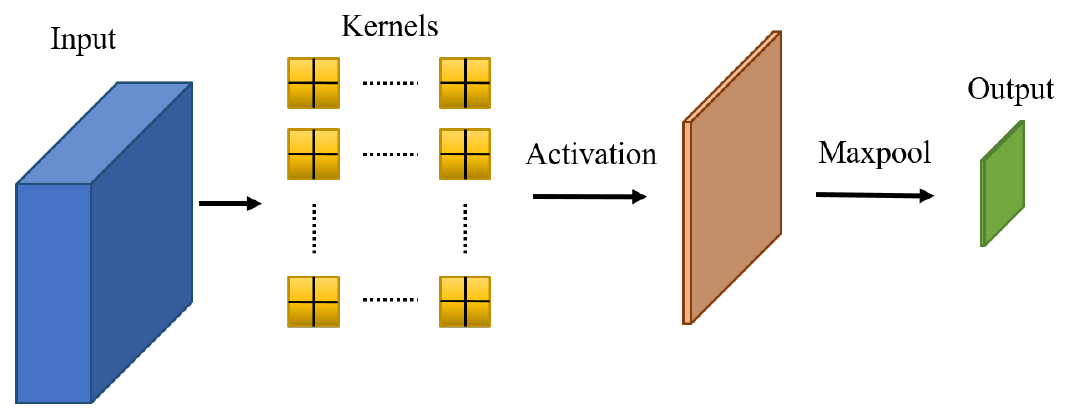}
    \caption{A typical single layer of CNN includes the convolution (see \eqref{eq:conv}) or cross-correlation operation (see \eqref{eq:crosscorr}) followed by the activation function and a maxpooling operation.}
    \label{fig:cnn_1layer}
\end{figure}
\begin{figure}[!htb]
\centering
 \includegraphics[scale=1.0]{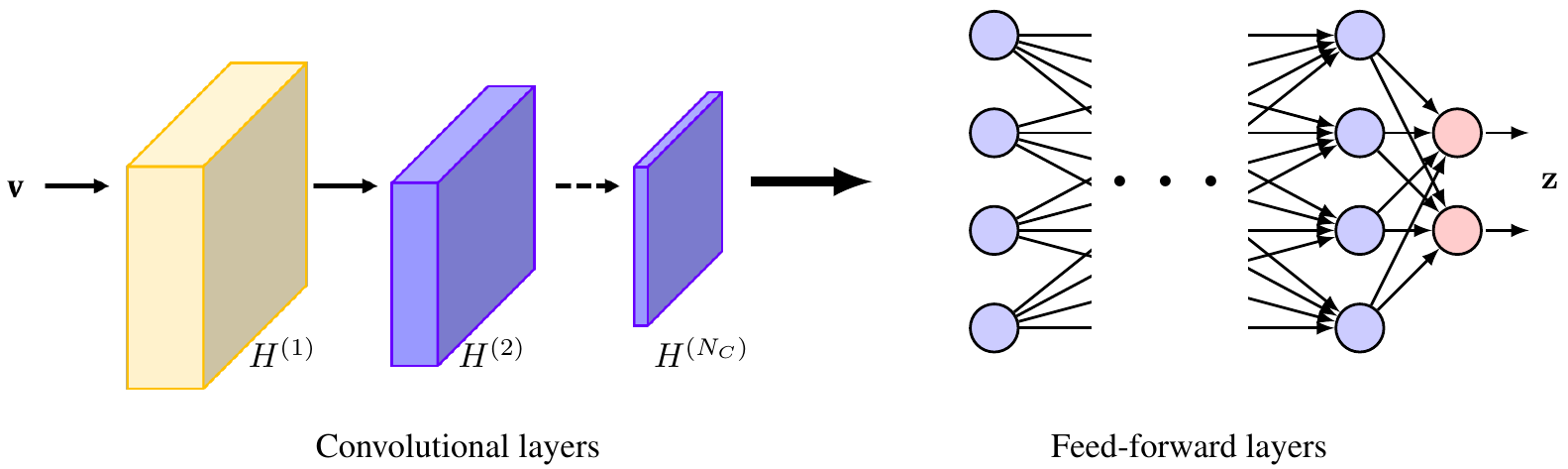}
\caption{A schematic of the convolutional neural network (CNN), where ${N_C}$ convolutional layers are followed by a few feed-forward layers to produce the output $\zm$ given the input $\vm$. } \label{fig:cnn}
\end{figure}

The convolutional neural network (CNN) has been developed with inspiration from the vision system at the primary visual cortex of human brain \cite{goodfellow2016deep}. 
In CNN, the convolution operation is performed for a two-dimensional input $\vm$ and a kernel $\boldsymbol{\Psi}$ as follows 
\begin{equation}\label{eq:conv}
    s_{ij} = \sum_{q}\sum_r v_{i-q,j-r} \Psi_{qr},
\end{equation} 
where $\sm$ is the two-dimensional output from a convolutional layer. During training of the CNN, the kernel $\boldsymbol{\Psi}$ is learned. 
Note that, zero-padding is required if the output $\sm$ and input $\vm$ are of the same length. Pytorch \cite{paszke2017automatic}, which is used for the numerical examples herein, however, performs the cross-correlation instead of the convolution given by
\begin{equation} \label{eq:crosscorr}
    s_{ij} = \sum_q \sum_r v_{i+q,j+r}\Psi_{qr}
\end{equation}
for a two-dimensional input $\vm$, which uses a mirror image of the kernel in \eqref{eq:conv}. Figure \ref{fig:conv} illustrates the working of this procedure, where blue shaded elements of $\vm$ are multiplied by $\boldsymbol\Psi$ to get the red shaded element in $\sm$. A maxpooling operation often follows a convolution operation in which the output from the convolution layer is downsampled using a max function over a window. However, in time histories if the length of both input and output remains same the maxpooling has limited use \cite{zhang2019physics} and can be omitted. These steps inside a single convolutional layer are shown in Figure \ref{fig:cnn_1layer}. Note that a kernel with size smaller than the size of the input produces sparse connectivity in the network. This creates a sharing of the parameters and helps in avoiding over-fitting. 
A typical implementation of the CNN often uses a few convolutional layers followed by feed-forward layers as shown in Figure \ref{fig:cnn}.  
Note that for $N$ input of images to the CNN the input has a size $N\times C\times H \times W$, where $C$ is the number of channels (\textit{e.g.}, R-G-B); $H$ is the height; and $W$ is the width of the images.  For $N$ input of time-histories, which is used here, the input has a size $N\times C\times L$, where $L$ is the length of time-histories. Here, the response and the uncertain parameters are used as different channels in the input. Further, the number of neurons in the feed-forward layers are assumed to be same as $L$ in this paper. 


\subsubsection{Training of a neural network}
The parameters $\thetaa$ of a neural network, which can contain the weights and biases $\Big\{ \{\Wm_i\}_{i=0}^{N_H}, \{\bmm_i\}_{i=0}^{N_H} \Big\}$ and the set of kernels $\Big\{\{\Psi_i\}_{i=1}^{N^{(C)}}\Big\}$, are learned from the data by minimizing the mismatch between the prediction from the network and the training dataset. In this paper, the Mean Squared Error (MSE) given by
\begin{equation}
J=\frac{1}{\Ntr}\sum_{i=1}^{\Ntr} \Big(\zm_i - \pinn(\vm_i;\thetaa) \Big)^2
\end{equation}
is used as the cost function for the optimization problem. A regularization term can also be added to obtain sparsely connected network. 
Stochastic Gradient Descent (SGD) is commonly used to update the parameters during the optimization as follows
\begin{equation}
    \begin{split}
\thetaa_{k+1} &\leftarrow \thetaa_{k} - \eta_k\frac{\partial J}{ \partial \thetaa_{k}},\\
\end{split}
\end{equation}
where $\eta_k$ is the step size at the $k$th iteration, also known as the learning rate and back propagation is used to estimate the derivatives $\frac{\partial J}{ \partial \thetaa^{(k)}}$ \cite{goodfellow2016deep,higham2018deep}. In this paper, a modified version of this standard SGD, namely, the Adaptive Moment (Adam) \cite{kingma2014adam,de2019topology,de2019bi} is used. In this algorithm, historical information of the gradients is used to retard movements in the direction of large historical gradients. A brief description of the Adam algorithm is provided in Appendix \ref{sec:adam}.


\textbf{Remark:} The universal approximation theorem  \cite{cybenko1989approximation,hornik1989multilayer,hornik1990universal,hesthaven2018non} guarantees that MLP with at least one hidden layer and enough hidden neurons with differentiable activation functions is able to approximate any continuous function up to a given accuracy level. In practice, the limitations are two-fold --- (i) the training algorithm used to optimize the parameters of the network might be unable to find the optimal values; and (ii) required number of neurons in the hidden layer may be quite large. Hence, multiple hidden layers are used in practical applications.


\section{Proposed Methodology}
In this section, the proposed approach for the response calculation of an uncertain locally nonlinear dynamical system using neural networks is discussed first. Then, the datasets used in training and validation for the numerical examples are described. 


\subsection{Response Calculation of Locally Nonlinear Uncertain Dynamical System}
Consider a locally nonlinear dynamical system with governing differential equation in state-space form given by
\begin{equation}\label{eq:state-space}
\begin{split}
\dot{\Xm}(t;\xii) &= \Am \Xm(t;\xii) + \Bm \wm(t) + \Lm \gm\left(\Xmb(t;\xii);\xii\right), \quad \Xm(0) = \xm_0;\\
\Ym(t;\xii) &= \Cm \Xm(t;\xii) + \Dm \wm(t) + \Em\gm\left(\Xmb(t;\xii);\xii\right) , \\
\end{split}
\end{equation}
where $\Xm(t;\xii)\in\mathbb{R}^{n\times 1}$ is the state vector; $\Am\in \mathbb{R}^{n\times n}$ is the state matrix; $\wm(t)\in\mathbb{R}^{\nw\times 1}$ is the external force vector; $\Bm\in\mathbb{R}^{n\times \nw}$ is the influence matrix for $\wm(t)$;  $\gm(\cdot;\cdot)\in\mathbb{R}^{n\times \ngm}$ is a nonlinear function of a subset of the state, (\textit{i.e.}, $\Xmb(t;\xii)=\Gm\Xm(t;\xii)$ with $\Gm\in\mathbb{R}^{\ngm\times n}$ and $\ngm\ll n$) and the uncertain variable $\xii$ with known probability distribution; $\Lm\in\mathbb{R}^{n\times \ngm}$ is the influence matrix for the nonlinear function $\gm(\cdot,\cdot)$; and $\xm_0$ is the initial state vector. The output is denoted as $\Ym(t;\xii)\in\mathbb{R}^{\ny\times 1}$. The output influence matrices are $\Cm\in\mathbb{R}^{\ny\times n}$, $\Dm\in\mathbb{R}^{\ny\times \nw}$, and $\Em \in \mathbb{R}^{\ny\times \ngm}$ for the state vector $\Xm(t;\xii)$, external force $\wm(t)$, and the uncertain and possibly nonlinear function $\gm(\cdot;\cdot)$, respectively. For example, consider a multi-degree of freedom nonlinear mass-spring-damper system with governing equation
\begin{equation}
\Mm\ddot{\um}(t;\xii) + \Cms \dot{\um}(t;\xii) + \Km \um(t;\xii) + {\Lms} \gm_\mathrm{s}\left(\um(t;\xii),\dot{\um}(t;\xii);\xii\right) = \wm(t),
\end{equation}
where $\um(t;\xii)$ is the displacement vector; $\Mm\in\mathbb{R}^{m\times m}$ is the mass matrix; $\Cms\in\mathbb{R}^{m\times m}$ is the damping matrix;  $\Km\in\mathbb{R}^{m\times m}$ is the stiffness matrix; ${\Lms}\in \mathbb{R}^{m\times {n}_\mathrm{g_\mathrm{s}}}$ is the influence matrix of the nonlinear and uncertain vector $\gm_\mathrm{s}(\cdot,\cdot;\cdot)\in\mathbb{R}^{{n}_\mathrm{g_\mathrm{s}}\times 1}$. The state-space matrices for this system are as follows
\begin{equation}
\begin{split}
\Xm(t;\xii) & = \left\{\begin{array}{c}
\um(t;\xii)\\
\dot{\um}(t;\xii)\\
\end{array}\right\}, \quad
\Am = \left[ \begin{array}{c c}
\mathbf{0} & \mathbf{I}\\
-\Mm^{-1}\Km & -\Mm^{-1}\Cms\\
\end{array}\right],\quad
\Bm = \left\{
\begin{array}{c}
\mathbf{0}\\
\Mm^{-1}\mathbf{1}\\
\end{array}
\right\},\quad
\Lm = \left\{
\begin{array}{c}
\mathbf{0}\\
-\Mm^{-1}{{\Lms}}\\
\end{array}
\right\},
\end{split}
\end{equation}
where $\mathbf{I}$ is the identity matrix; $\mathbf{0}$ is a matrix with all entries as zeros; and $\mathbf{1}$ is a matrix with all entries as ones.
A deterministic nominal linear dynamical system corresponding to this uncertain nonlinear dynamical system can be given by 
\begin{equation}\label{eq:detresponse}
\begin{split}
\dot{\xm}(t) &= \Am \xm(t) + \Bm \wm(t), \quad \xm(0)=\xm_0;\\
\ym(t) &= \Cm \xm(t) + \Dm \wm(t),
\end{split}
\end{equation}
where $\xm(t)$ and $\ym(t)$ are state and output of the nominal dynamical system, respectively.

\begin{figure}[h]\centering
\includegraphics[scale=1]{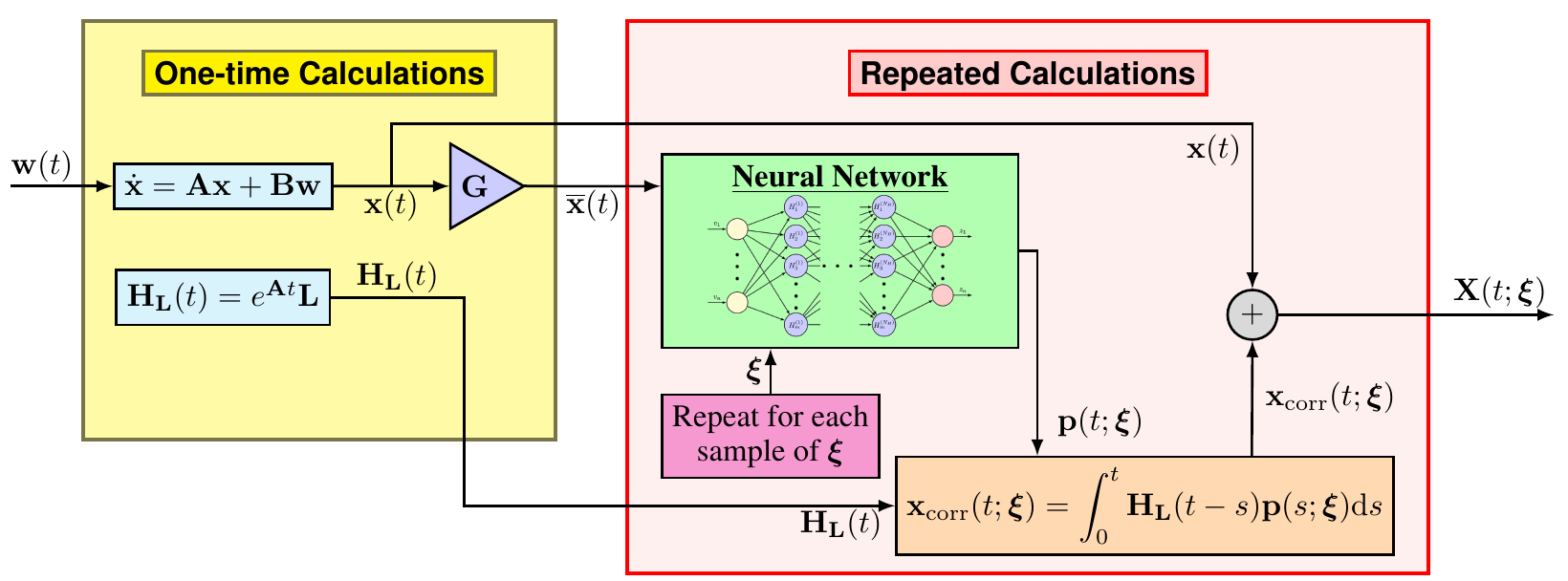}
	\caption{Implementation of the proposed approach for efficient uncertainty quantification showing one time calculation and repeated calculation components.}\label{fig:nn_schem}
\end{figure}

The response of the original uncertain locally nonlinear dynamical system is expressed, next, as summation of the response of the nominal linear system $\xm(t)$ from \eqref{eq:detresponse} and a correction term $\xnl(t;\xii)$ due to the nonlinearity and uncertainty present in the system, \textit{i.e.},
\begin{equation}
\Xm(t;\xii) = \xm(t)+\xnl(t;\xii).
\end{equation}
The response of the nominal linear system can be estimated using 
\begin{equation}\label{eq:linresp}
\begin{split}
\xm(t) &= \exp(\Am t) \xm_0 + \int_0^t \Hb(t-\tau)\wm(\tau)\drm\tau;\\
\ym(t) &= \Cm \exp(\Am t) \xm_0 + \int_0^t \Cm\Hb(t-\tau)\wm(\tau)\drm\tau + \Dm \wm(t),\\
\end{split}
\end{equation}
where the impulse response function $\Hb(t) = \exp(\Am t) \Bm$. Note that the convolution integral can be efficiently evaluated using the Fast Fourier Transform (FFT).
However, the most challenging part is to estimate $\xnl(t)$, which can be similarly written as
\begin{equation}\label{eq:xnl}
\begin{split}
\xnl(t;\xii) &= \int_0^t \Hl (t-\tau) \gm\left(\Xmb(\tau;\xii);\xii\right) \drm \tau;\\
\Ym(t;\xii) &= \ym(t) +  \underbrace{\int_0^t \Cm\Hl (t-\tau) \gm\left(\Xmb(\tau;\xii);\xii\right) \drm \tau+ \Em \gm\left(\Xmb(\tau;\xii);\xii\right)}_{=\ym_\mathrm{corr}(t;\xii)},
\end{split}
\end{equation}
where the impulse response function $\Hl(t) = \exp(\Am t ) \Lm$ is due to the nonlinear function $\gm(\cdot,\cdot)$ and $\ym_\mathrm{corr}(t;\xii)$ is the contribution to the output from $\gm(\cdot,\cdot)$. Here, $\xnl(t;\xii)$ implicitly depends on $\Xm(t;\xii)$ as can be seen in the solution of \eqref{eq:xnl}. An iterative and complex approach to solve \eqref{eq:xnl} using FFT and Newton's method was proposed in \cite{wojtkiewicz2011efficient} but requires $\partial \gm/\partial \xm$ and an efficient breakup of the convolution sum to achieve a computational speedup. In this paper, instead, neural networks are employed to model a pseudoforce defined as
\begin{equation}\label{eq:p}
\ppm(t;\xii) \approx \gm\left( \Xmb(t;\xii);\xii \right). 
\end{equation}
Figure \ref{fig:nn_schem} shows the proposed approach with one-time calculation to estimate the nominal response $\xm(t)$ and repeated calculation using the neural network to estimate the response $\xnl(t;\xii)$, where $\xo(t) = \Gm \xm(t)$ is also used as the input for the neural network. 
Note that, $\ppm(t;\xii)$ is also the solution of a nonlinear Volterra integral equation of the second kind written in a non-standard form given by 
\begin{equation}\label{eq:nvie}
    \begin{split}
        &\ppm(t;\xii) - \gm\left(\xmb(t)+\int_0^t \Hl (t-\tau) \ppm\left(\tau;\xii\right) \drm \tau;\xii\right)=\mathbf{0}.\\
    \end{split}
\end{equation}
The solution of \eqref{eq:nvie}, however, requires an iterative strategy due to the nonlinearity \cite{wojtkiewicz2011efficient,de2018computationally,de2017efficient}. 

\subsection{Datasets used in Training of the Neural Networks} \label{sec:datasets}
To train these neural networks, two datasets, namely, the training dataset $\Dtr=\Bigg\{\Big\{\ppm^{}\left(t_i;\xii_j^{(\mathrm{tr})}\right)\Big\}_{i=1}^{\nt},\xii_j^{(\mathrm{tr})}\Bigg\}_{j=1}^{\Ntr}$ and the validation dataset $\Dval=\Bigg\{\Big\{\ppm^{}\left(t_i;\xii_j^{(\mathrm{val})}\right)\Big\}_{i=1}^{\nt},\xii_j\Bigg\}_{j=1}^{\Nval}$ are generated by solving \eqref{eq:nvie} at time instances $\{t_i\}_{i=1}^{\nt}$ for random $\Ntr$ and $\Nval$ realizations of $\xii$, respectively, where $\{\xii_j\}_{j=1}^{\Ntr}$ and $\{\xii_j\}_{j=1}^{\Nval}$ do not overlap. 
For neural networks, $\xii$ and $\xo(t)$ are used for input $\vm$ and $\ppm(t;\xii)$ is used as the output $\zm$ (see \eqref{eq:nn}, \eqref{eq:resnet}, and Figure \ref{fig:cnn}). 
The training dataset $\Dtr$ is used to estimate the gradients and the validation dataset $\Dval$ is used to monitor the validation Root Mean Squared  Error (RMSE) defined as 
\begin{equation}\label{eq:val_rmse} 
    \text{RMSE} = \frac{\sum_{i=1}^{N_\mathrm{val}}\Big\lVert \ppm^{}\left(t;\xii_j^{(\mathrm{val})}\right) - \pnn^{}\left(t;\xii_j^{(\mathrm{val})}\right) \Big\rVert_2}{\sum_{i=1}^{N_\mathrm{val}}\Big\lVert \ppm\left(t;\xii_j^{(\mathrm{val})}\right) \Big \rVert_2},
\end{equation}
where $\ppm\left(t;\xii_j^{(\mathrm{val})}\right)$ is the prediction using the neural network and $\lVert \cdot \rVert_2$ is the Euclidean norm. 
For FNN and ResNet architectures, an iterative procedure is followed to select the number of hidden layers $N_H$ and the number of neurons per hidden layer $m$ \cite{de2020transfer}, where $m$ is increased gradually up to a maximum while a validation error is monitored. The number of hidden layers $N_H$ is increased by one if a pre-chosen maximum neurons per layer is reached. The final configuration is chosen that corresponds to the smallest validation error. For CNN, a similar procedure with the number of convolution layers is followed. 
The training of these neural networks require a few hours on a modern desktop. 
Once trained these networks produce inexpensive but accurate prediction of the response of the locally nonlinear dynamical system under uncertainty as the next three numerical examples show. The same trained network can be used even when the quantity of interest depends on different responses. Further, they can be also used for other applications such as design under uncertainty, sensitivity analysis, and so on. 


\section{Numerical Examples} \label{sec:ex}
Three numerical examples utilizing structures with increasing number of DOF are used in this section to illustrate the proposed approach. PyTorch \cite{paszke2017automatic} is used to implement the neural networks for the examples. 
For brevity, the dependence of the quantities on $t$ and $\xii$ are omitted in this section. In the examples, the accuracy of the estimates is measured using the Root Mean Squared Error (RMSE) given by $\frac{\lVert \widehat{\ym} - \ym \rVert_2}{\lVert \ym \rVert_2}$, where $\widehat{\ym}$ is the estimated quantity and $\ym$ is the true response.

\subsection{Example I: Two Degree-of-freedom Nonlinear Spring-Mass-Damper Model}
\begin{figure}[!htb]
  \begin{center}
     \includegraphics[scale=1.0]{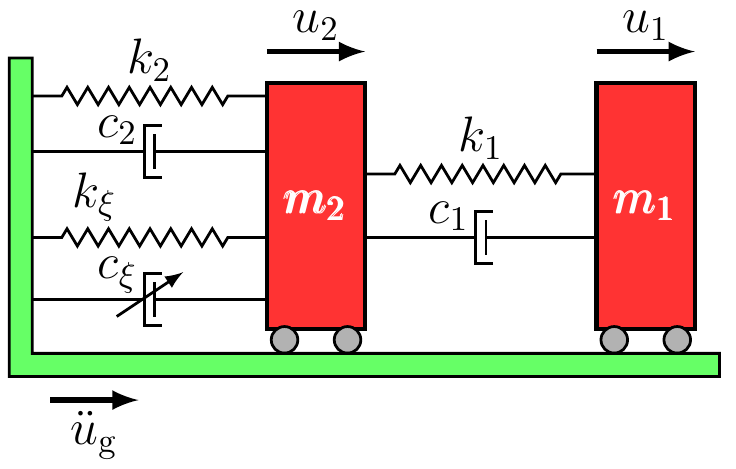}
  \end{center}
  \caption{Two DOF spring-mass-damper model with uncertain stiffness $k_\xi$ and nonlinear damping $c_\xi$ used in Example I.} \label{fig:2dof}
\end{figure}

A two DOF spring-mass-damper model with nonlinear damping is used in this example (see Figure \ref{fig:2dof}). 
The equations of motion for this model are given by
\begin{equation}
\begin{split}
    &m_1 \ddot{u}_1 + c_1(\dot{u}_1 - \dot{u}_2) + k_1 (u_1 - u_2) = -m_1 \ddot{u}_g;\\
    &m_2 \ddot{u}_2 + c_1(\dot{u}_2 - \dot{u}_1) + c_2 \dot{u}_2  +  c_\xi(\dot{u}_2)^3 + k_1 (u_2 - u_1) + (k_2 + k_\xi)u_2 = -m_2 \ddot{u}_g,\\
\end{split}
\end{equation}
where the state vector is $\Xm = [u_1\quad u_2 \quad \dot{u}_1 \quad \dot{u}_2]^T$ and $\ddot{u}_\mathrm{g}$ is the ground acceleration. To generate the training and validation datasets, the following specifications are used: $m_1=29,485$ kg, $m_2=6,800$ kg, $k_1=11,912$ kN/m, $k_2=250$ kN/m, $c_1=23.71$ kN$\cdot$s/m, and $c_2=4$ kN$\cdot$s/m, where the parameters are selected from \cite{ramallo2002smart}. The uncertainty is assumed in the stiffness $k_\xi$ and damping coefficient $c_\xi$. The probability distributions for these two parameters are shown in Table \ref{tab:ExI_param}. The base excitation used is a stationary filtered white noise from a Kanai-Tajimi filter \cite{lin1987evolutionary} that has a spectral density
\begin{equation}\label{eq:kanai-tajimi}
    S_{\ddot{u}_g\ddot{u}_g}(\omega) = \frac{S_0(4\zeta_g^2\omega_g^2\omega^2+\omega_g^4)}{(\omega^2-\omega_g^2)^2+4\zeta_g^2\omega_g^2\omega^2},
\end{equation}
where $\omega_g=17$ rad/s and $\zeta_g=0.3$ are chosen from \cite{ramallo2002smart}. The spectral intensity $S_0$ is given by
\begin{equation}
    S_0=\sigma_w^2\frac{0.03\zeta_g}{\pi \omega_g(4\zeta_g^2+1)}g^2,
\end{equation}
where $g$ is the gravitational acceleration and $\sigma_w=2$ is selected to have a pronounced nonlinearity in the system response. In the state-space form of \eqref{eq:state-space}, $\Lm=[0\quad 0\quad 0\quad -1/m_2]^T$ and $g({\Xmb})=k_\xi u_2 + c_\xi (\dot{u}_2)^3$, where $\Xmb = [u_2\quad \dot{u}_2]^T$. 
\begin{center}

\begin{threeparttable}[!htb]
\caption{Probability distribution of the uncertain parameters in Example I. } \label{tab:ExI_param} 
\begin{tabular}{c c c c} 
\hline 
\Tstrut
Parameter & Distribution & Mean & Std. Dev. \\ [0.5ex] 
\hline 
\Tstrut
$k_\xi$ & Truncated Gaussian$^*$ & 40 kN/m & 10 kN/m \\ 
$c_\xi$ & Lognormal & 75 kN$\cdot$(s/m)$^{1/3}$ & 20 kN$\cdot$(s/m)$^{1/3}$ \\ [1ex] 
\hline 
\end{tabular}
\begin{tablenotes}
\item[$^*$] Truncated below at zero. 
\end{tablenotes}
\end{threeparttable}
\end{center}

\begin{figure}[!htb]
    \centering
    \includegraphics[scale=0.3]{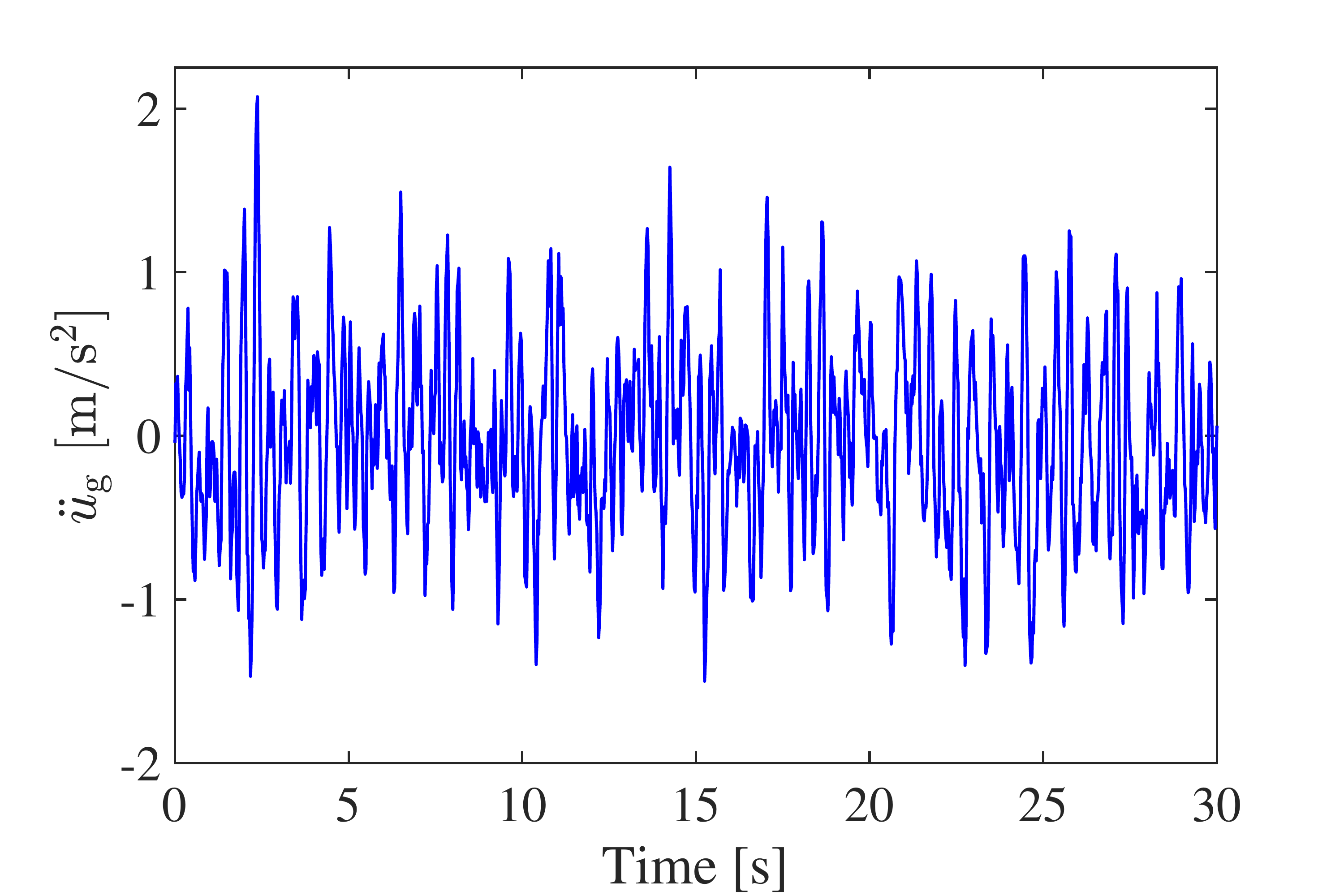}
    \caption{Base excitation generated using a Kanai-Tajimi filter in \eqref{eq:kanai-tajimi} and used in Example I.}
    \label{fig:xgdd}
\end{figure}
\subsubsection{Results} 

\begin{figure}[!htb]
    \centering
    \includegraphics[scale=0.3]{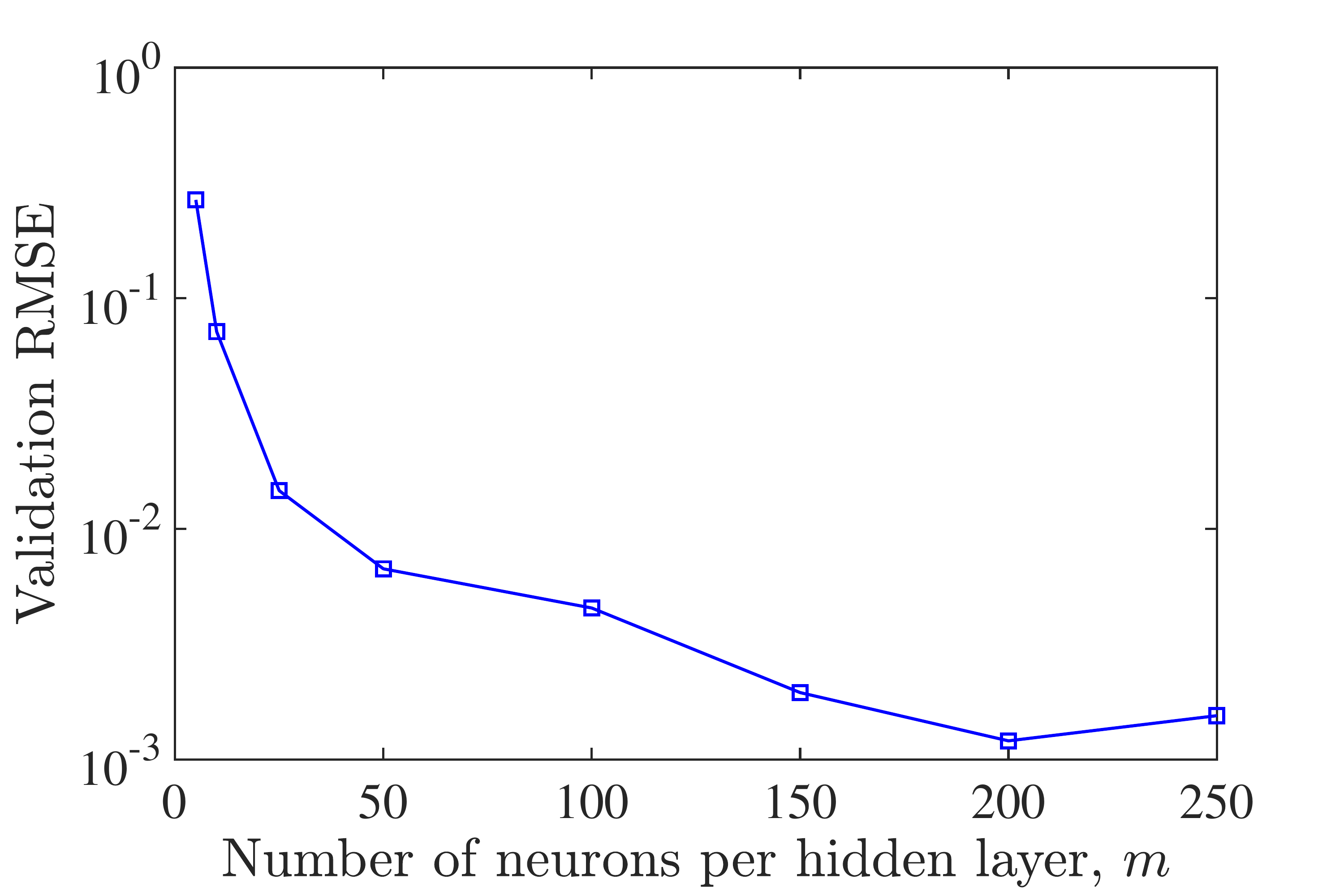}
    \caption{The validation RMSE decreases as more neurons are added to four hidden layers in FNN in Example I.}
    \label{fig:neurons}
\end{figure}

\begin{figure}[!htb]
    \centering
    \begin{subfigure}[t]{0.48\textwidth}
    \centering
    \includegraphics[scale=0.3]{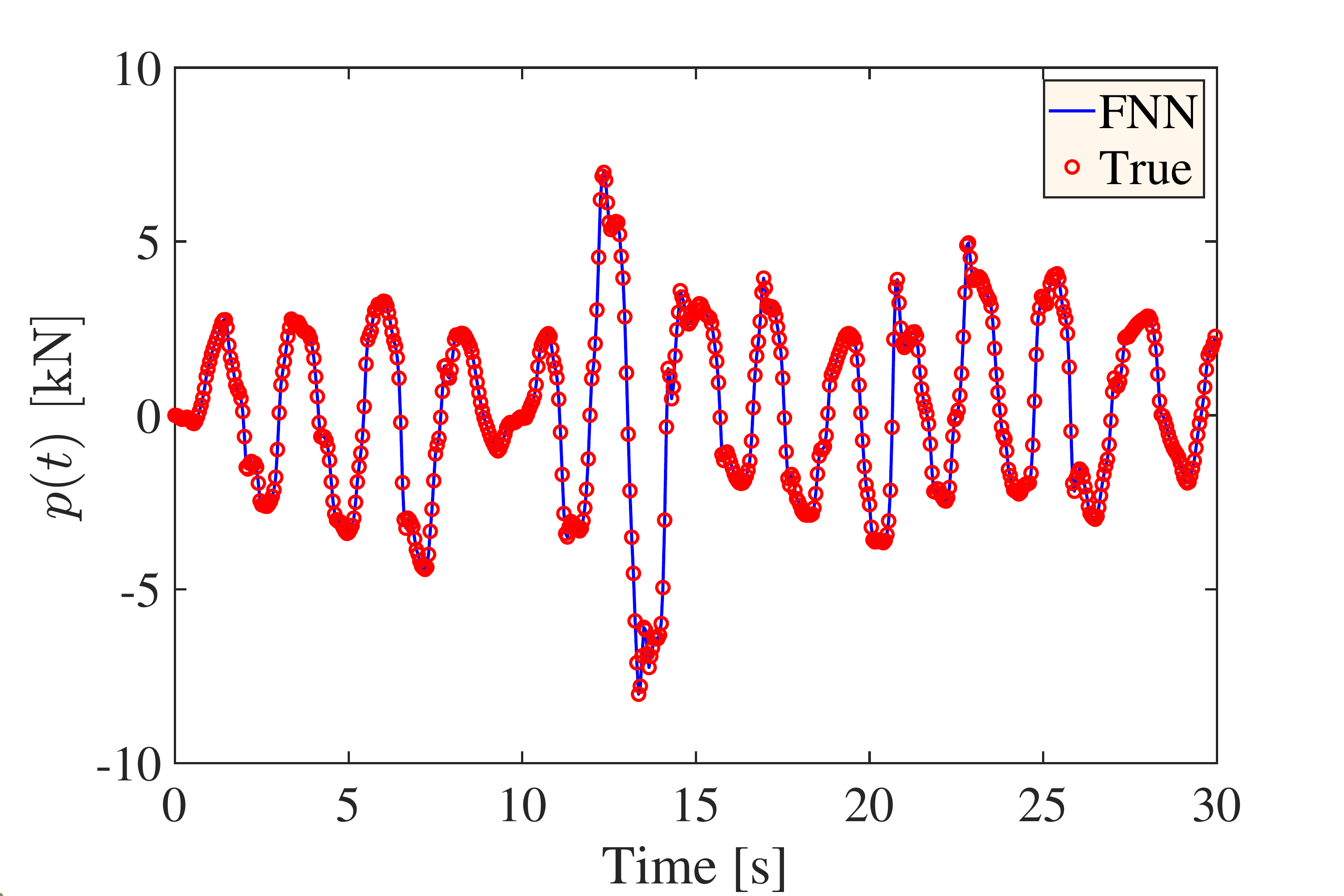}
    \caption{Prediction of $p(t)$ using FNN}
    \end{subfigure}
    \hfill
    \begin{subfigure}[t]{0.48\textwidth}
    \centering
    \includegraphics[scale=0.3]{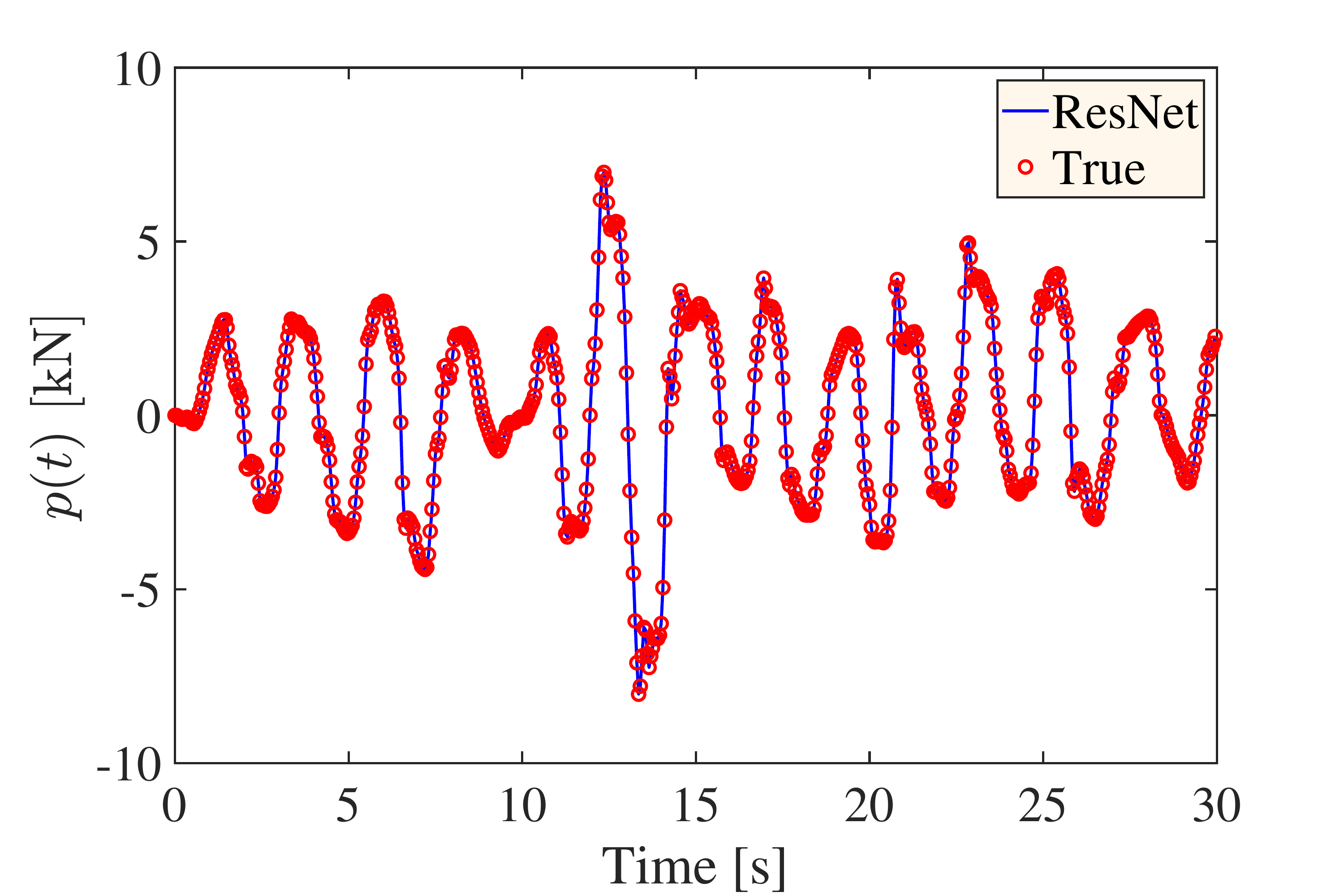}
    \caption{Prediction of $p(t)$ using ResNet}
    \end{subfigure}\\
    \begin{subfigure}[t]{0.48\textwidth}
    \centering
    \includegraphics[scale=0.3]{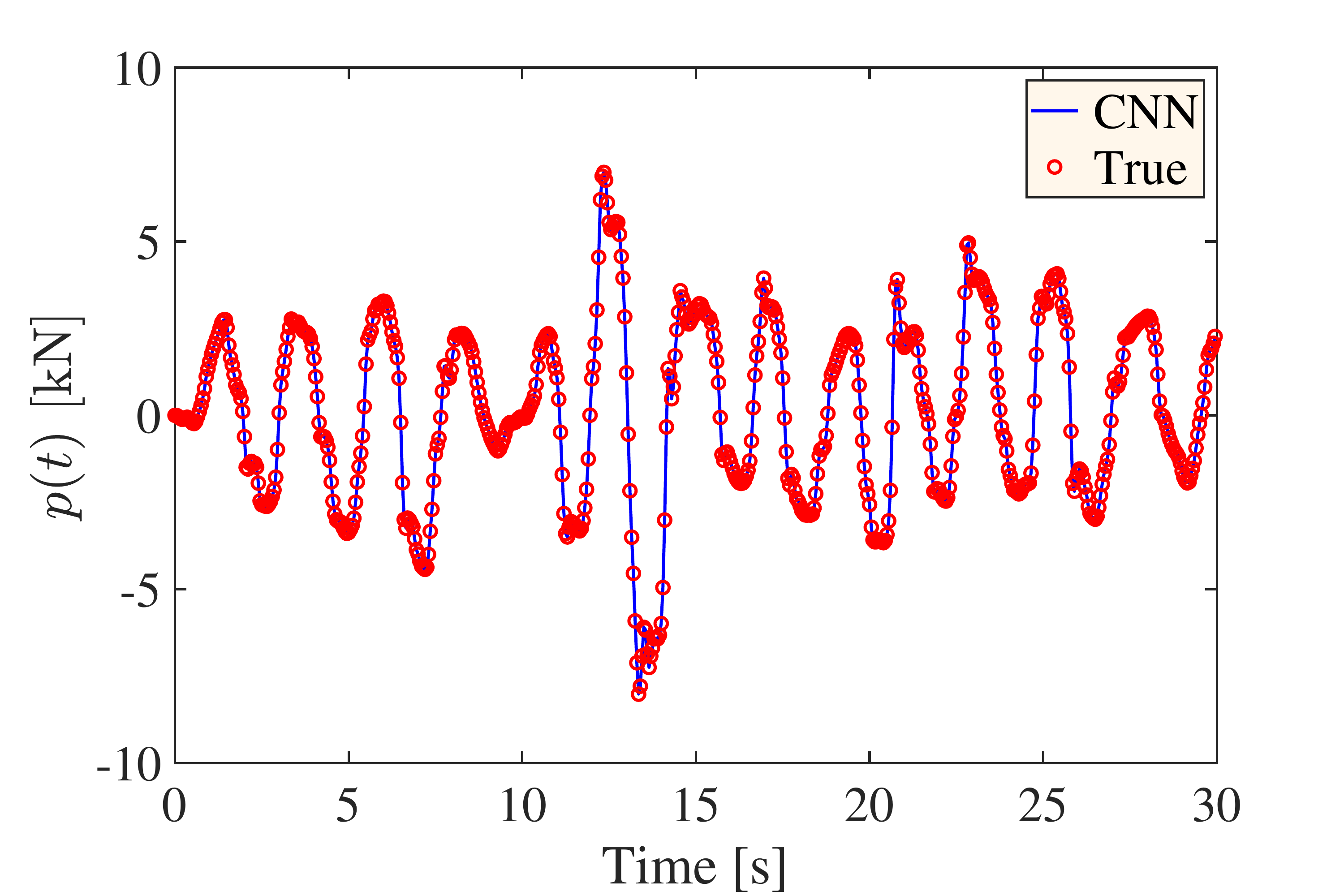}
    \caption{Prediction of $p(t)$ using CNN}
    \end{subfigure}
    \caption{Comparison of predictions from three different neural network architectures for a realization of the uncertainty in the validation dataset $\Dval$ in Example I. The true solution is obtained by solving \eqref{eq:nvie}.}
    \label{fig:2dof_p_valid}
\end{figure}

The neural networks are used to model the uncertain nonlinear term $p(t;\xii)=g(\Xmb)=k_\xi u_2+c_\xi (\dot{u}_2^3)$. The training dataset $\Dtr=\Bigg\{\Big\{p\left(t_i;\xii_j^{(\mathrm{tr})}\right)\Big\}_{i=1}^{\nt},\xii_j^{(\mathrm{tr})}\Bigg\}_{j=1}^{\Ntr}$ is generated using $\Ntr=250$ random samples of the uncertain parameters drawn from their respective probability distributions given in Table \ref{tab:ExI_param} and with a 20 Hz temporal sampling rate. For validation dataset $\Dval=\Bigg\{\Big\{p\left(t_i;\xii_j^{(\mathrm{val})}\right)\Big\}_{i=1}^{\nt},\xii_j^{(\mathrm{val})}\Bigg\}_{j=1}^{\Nval}$, separate $\Nval=50$ random samples are used. 
To select the number of neurons per layer $m$ and total number of hidden layers $N_H$ a procedure described in Section \ref{sec:datasets} is followed. Figure \ref{fig:neurons} shows $4$ layers and $200$ neurons per layer produces the smallest validation error with FNN architecture. The activation function is chosen as the sigmoid function $\sigma_\mathrm{sigm}(\cdot)$ (see \eqref{eq:act1}) as it gives the smallest validation RMSE. ResNet uses a similar configuration with a residual connection between the first and third layer. 
For CNN, a similar procedure is followed and $N_C=3$ one-dimensional convolution layers with kernels of length three followed by two feed-forward layers with $n_t$ neurons each are used. The activation function for the convolution layers are chosen as the sigmoid function $\sigma_\mathrm{sigm}(\cdot)$, whereas the feed-forward layers use the ELU activation $\sigma_{\!_\mathrm{ELU}}(\cdot)$ (see \eqref{eq:act2}). 
Adam algorithm briefly described in Appendix \ref{sec:adam} is used to train these networks with a learning rate of $10^{-3}$, which is gradually halved every 2000 iterations for training of FNN and ResNet but halved every 500 iterations for training of CNN subjected to a maximum iteration of 10000. This schedule of learning rate produces converged result from the optimization. The training of FNN and ResNet architectures took approximately 4 hrs. whereas the training of CNN took approximately 6 hrs.
Note that these trainings are performed on CPU (central processing unit). However, by performing them on GPU (graphics processing unit) the training time can be significantly reduced. At the conclusion of training, the trained network is chosen as the network that produces the smallest validation RMSE. This is equivalent to an early stopping criterion \cite{bengio2012practical} commonly used in machine learning applications with infinite patience and subject to a maximum iteration count. 
Table \ref{tab:2_val_rmse} shows the validation RMSE using each of the three architectures, where CNN produces the smallest error as it implements a sharing of the network parameters. Figure \ref{fig:2dof_p_valid} shows the estimated $p(t;\xii)$ using these architectures for one realization of the uncertain parameters in the validation dataset $\Dval$. 

\begin{table}[!htb]
\caption{Validation RMSE for $g(\Xmb)$ using three different architectures for the neural networks in Example I. } 
\centering 
\begin{tabular}{c c c c} 
\hline 
\Tstrut
NN architecture & Validation RMSE \\ [0.5ex] 
\hline 
\Tstrut
 FNN & $1.9506\times10^{-3}$ \\ 
 ResNet & $3.2809\times10^{-3}$ \\
 CNN & $1.2765\times10^{-3}$ \\ [1ex] 
\hline 
\end{tabular}
\label{tab:2_val_rmse} 
\end{table}

\begin{table}[!htb]
\caption{RMSE of mean and standard deviation of the base displacement $u_2$ using three different architectures for the neural networks in Example I. } 
\centering 
\begin{tabular}{c c c c} 
\hline 
\Tstrut
NN architecture & RMSE of mean & RMSE of std dev \\ [0.5ex] 
\hline 
\Tstrut
 FNN & $1.1369\times10^{-3}$ & $8.5926\times10^{-3}$\\ 
 ResNet & $1.2870\times10^{-3}$ & $8.7527\times10^{-3}$\\
 CNN & $1.0753\times10^{-3}$ & $8.1803\times10^{-3}$\\ [1ex] 
\hline 
\end{tabular}
\label{tab:2_pred_rmse} 
\end{table}

\begin{figure}[!htb]
    \centering
    \begin{subfigure}[t]{0.48\textwidth}
    \centering
    \includegraphics[scale=0.9]{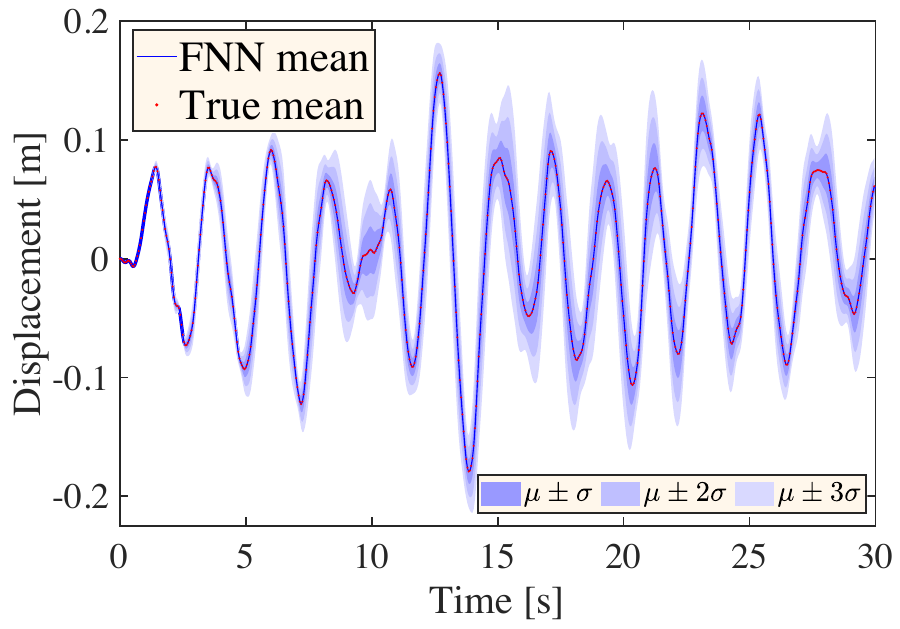}
    
    \caption{Prediction of mean and standard deviation of $u_2$ using FNN}
    \end{subfigure}
     \hfill
    \begin{subfigure}[t]{0.48\textwidth}
    \centering
    \includegraphics[scale=0.9]{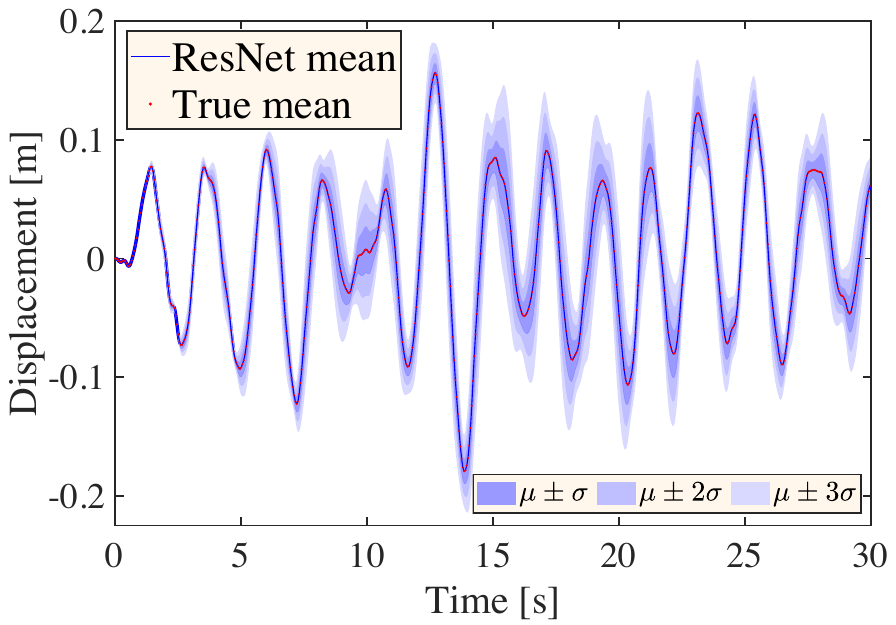}
    \caption{Prediction of mean and standard deviation of $u_2$ using ResNet}
    \end{subfigure}
    \begin{subfigure}[t]{0.48\textwidth}
    \centering
    \includegraphics[scale=0.9]{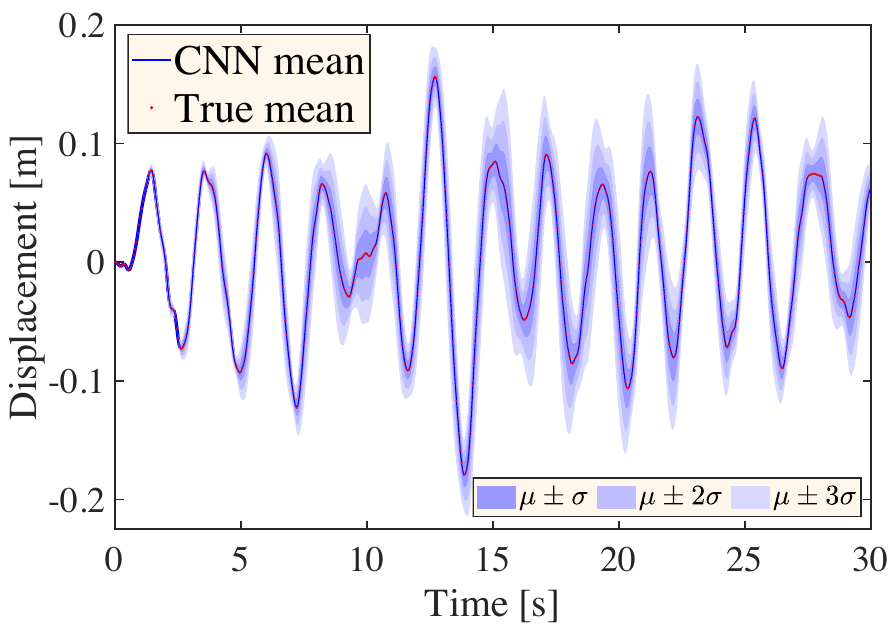}
    \caption{Prediction of mean and standard deviation of $u_2$ using CNN}
    \end{subfigure}
    \caption{Prediction of the mean and standard deviations of $u_2$ from three different neural network architectures using $10^5$ realizations of the uncertainty in Example I. The mean responses from the neural networks are compared with mean estimated from solving \eqref{eq:nvie} for $10^5$ realizations of the uncertainty. The shaded regions show combinations of mean $\mu$ and standard deviation $\sigma$ of the response.} 
    \label{fig:2dof_pred} 
\end{figure}

Next, the mean and standard deviation of the displacement $u_2$ of the base mass $m_2$ are estimated using the proposed approach with $N=10^5$ random samples. Figure \ref{fig:2dof_pred} shows the result and compares to the mean displacement obtained from solving 
Table \ref{tab:2_pred_rmse} reports the RMSE of the mean and standard deviation of the response $u_2$, which shows that the CNN architecture slightly outperforms the others. Also, the validation RMSE for the prediction of standard deviation is larger as it is more difficult to estimate this statistic. 
Once trained the FNN takes a total $3.94$ min. in for predicting responses for $10^5$ different realizations of the uncertainty. ResNet and CNN take $4.19$ min. and $1.82$ min., respectively, in total for $10^5$ evaluations. On the other hand, to solve \eqref{eq:nvie} using a computationally efficient method \cite{de2018computationally} takes $1.27$ hrs. in total for these evaluations. Note that \textsc{Matlab}'s \texttt{ode45}, a standard nonlinear solver, which does not have any associated one-time cost, however, takes $6.37$ s for one evaluation and hence it would take approximately 7 days for $10^5$ evaluations if this solver is used. The difference in computational cost is even more pronounced for complex structures as shown in the next two examples. 
Note that a desktop with octa-core Intel $i9-9900k ~@3.60$ GHz processor and 64 GB of RAM and running Wondows 10 is used to estimate the computation time.

\subsection{Example II: 11 Story Base Isolated Building}

\begin{figure}[!htb]
    \centering
    \begin{subfigure}[t]{0.45\textwidth}
    \centering
    \includegraphics[scale=0.5]{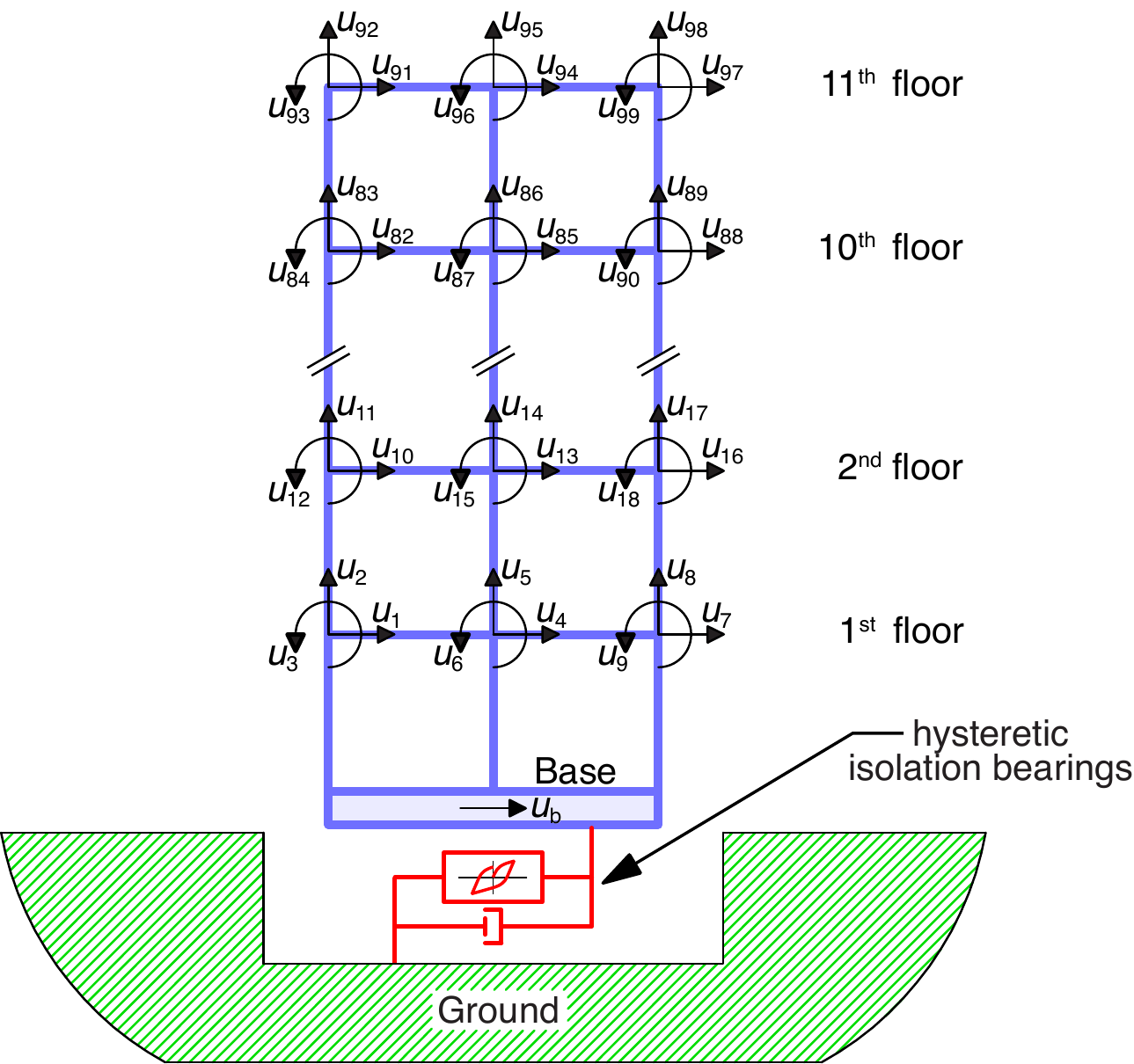}
    \caption{A 100 degree-of-freedom structure with a hysteretic base isolation layer.}\label{fig:100dofa}
    \end{subfigure}
    \hfill
    \begin{subfigure}[t]{0.5\textwidth}
    \centering
    \includegraphics[scale=1.0]{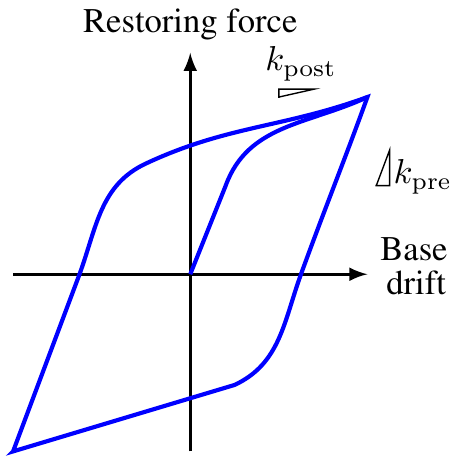}
\caption{Bouc-Wen model for hysteresis.}\label{fig:100dofb}
\end{subfigure}
    \caption{An 11-story 2-bay 100 degree-of-freedom structure with a hysteretic base isolation layer described using the Bouc-Wen hysteresis model is used in Example II. }
    \label{fig:100dof}
\end{figure}

In the second example, a 11-story 2-bay structural model with a hysteretic base isolation layer as shown in Figure \ref{fig:100dofa} is used. The base layer is assumed rigid in-plane and moving horizontally. The beams are modeled using consistent mass matrix and weights of the columns are neglected. The governing equations of this structure are given by 
\begin{equation}
\begin{split}
    &\Mm \ddot{\um}_\mathrm{s} + \Cms \dot{\um}_\mathrm{s} + \Km \um_\mathrm{s} = - \Mm \rr \ddot{u}_\mathrm{g} + \Cms \rr \dot{u}_\mathrm{b} + \Km \rr u_\mathrm{b};\\
    & m_\mathrm{b} \ddot{u}_\mathrm{b} + \left( c_\mathrm{b} + \rr^T \Cms \rr \right) \dot{u}_\mathrm{b} + \left( k_\mathrm{b} + \rr^T \Km \rr \right) {u}_\mathrm{b} + f_\mathrm{b} = -m_\mathrm{b}\ddot{u}_\mathrm{b} + \rr^T \Cms \rr \dot{\um}_\mathrm{s} + \rr^T \Km \rr \um_\mathrm{s},\\
\end{split}
\end{equation}
where $\Mm$ is the mass matrix, $\Cms$ is the damping matrix, and $\Km$ is the stiffness matrix of the superstructure; $\um$ is the displacement of the superstructure relative to the ground; $\ddot{u}_\mathrm{g}$ is the ground acceleration; and the influence vector for the ground acceleration is $\rr = [1,0,0,\dots,1,0,0]^T$, where the ones correspond to the horizontal displacement DOF. Note that the column weights are neglected but for the beams consistent mass matrix is used. The superstructure uses 33 nodes and three DOF per node. Hence, the combined structure with another DOF for the base layer has a total 100 DOF. 
The above equation can be converted to the state-space formulation in \eqref{eq:state-space} using $\Xm = [\um_\mathrm{s}^T\quad u_\mathrm{b} \quad \dot{\um}_\mathrm{s}^T \quad \dot{u}_\mathrm{b}]^T$. 
Rayleigh damping with 3\% damping ratios for the 1$^\mathrm{st}$ and 10$^\mathrm{th}$ superstructure modes is assumed. 
A record of the $1940$ El Centro earthquake measured at the N-S Imperial Valley Irrigation District substation with peak ground acceleration $0.348g$ is used as the ground acceleration $\ddot{u}_\mathrm{g}$. 

The nonlinearity is assumed in the restoring force of the hysteretic base layer described using the Bouc-Wen model (see Figure \ref{fig:100dofb}) \cite{wen1976method}, which gives the sum of the base layer restoring and damping forces as
\begin{equation}
    g(u_\mathrm{b},\dot{u}_\mathrm{b}) =
    f_\mathrm{b} = c_\mathrm{b}\dot{u}_\mathrm{b}+\kpost{u}_\mathrm{b}+\alpha z_\mathrm{hyst},
\end{equation}
where $c_\mathrm{b}$ is the base layer damping coefficient; $\kpre$ and $\kpost$ are pre- and post-yield stiffness, respectively; $\alpha = \Qy(1-r_k)$ is the peak of the nonelastic force; $\Qy$ is the yield force of the hysteretic layer; $r_k=\kpost/\kpre$ is the hardening ratio; and $z_\mathrm{hyst}$ is an auxiliary variable that describes the evolution of the hysteretic loop. The evolution of $z_\mathrm{hyst}$ is given by
\begin{equation}
    \dot{z}_\mathrm{hyst} = A \dot{u}_\mathrm{b} - \beta \dot{u}_\mathrm{b} |z_\mathrm{hyst}| - \gamma z_\mathrm{hyst} |\dot{u}_\mathrm{b}|,
\end{equation}
where $A=2\beta=2\gamma=\kpre/\Qy$ produces identical loading and unloading curves \cite{ramallo2002smart,ma2004parameter} with $z_\mathrm{hyst}$ in $[-1,1]$. 
Four parameters, namely, $c_\mathrm{b}$, $\kpost$, $\Qy$, and $r_k$ are assumed uncertain. The assumed probability distributions of these parameters are given in Table \ref{tab:ExII_param}. 

\begin{center}

\begin{threeparttable}[!htb]
\caption{Probability distribution of the uncertain parameters in Example II. } \label{tab:ExII_param} 
\begin{tabular}{c c c c} 
\hline 
\Tstrut
Parameter & Distribution & Mean & Std. Dev. \\ [0.5ex] 
\hline 
\Tstrut
$\kpost$ & Lognormal & 250 kN/m & 20 kN/m \\ 
$c_\mathrm{b}$ & Truncated Gaussian$^*$ & 3.5 kN$\cdot$s/m & 0.25 kN$\cdot$s/m \\
$\Qy$ (\%)$^\dagger$ & Uniform & 5.0 & 0.5774 \\
$r_k$ & Uniform & 0.1875 & 0.0361 \\ [1ex] 
\hline 
\end{tabular}
\begin{tablenotes}
\item[$^*$] Truncated below at zero. 
\item[$^\dagger$] in \% of the total weight of the structure.
\end{tablenotes}
\end{threeparttable}
\end{center}

\subsubsection{Results}

\begin{figure}[!htb]
    \centering
    \begin{subfigure}[t]{0.48\textwidth}
    \centering
    \includegraphics[scale=0.3]{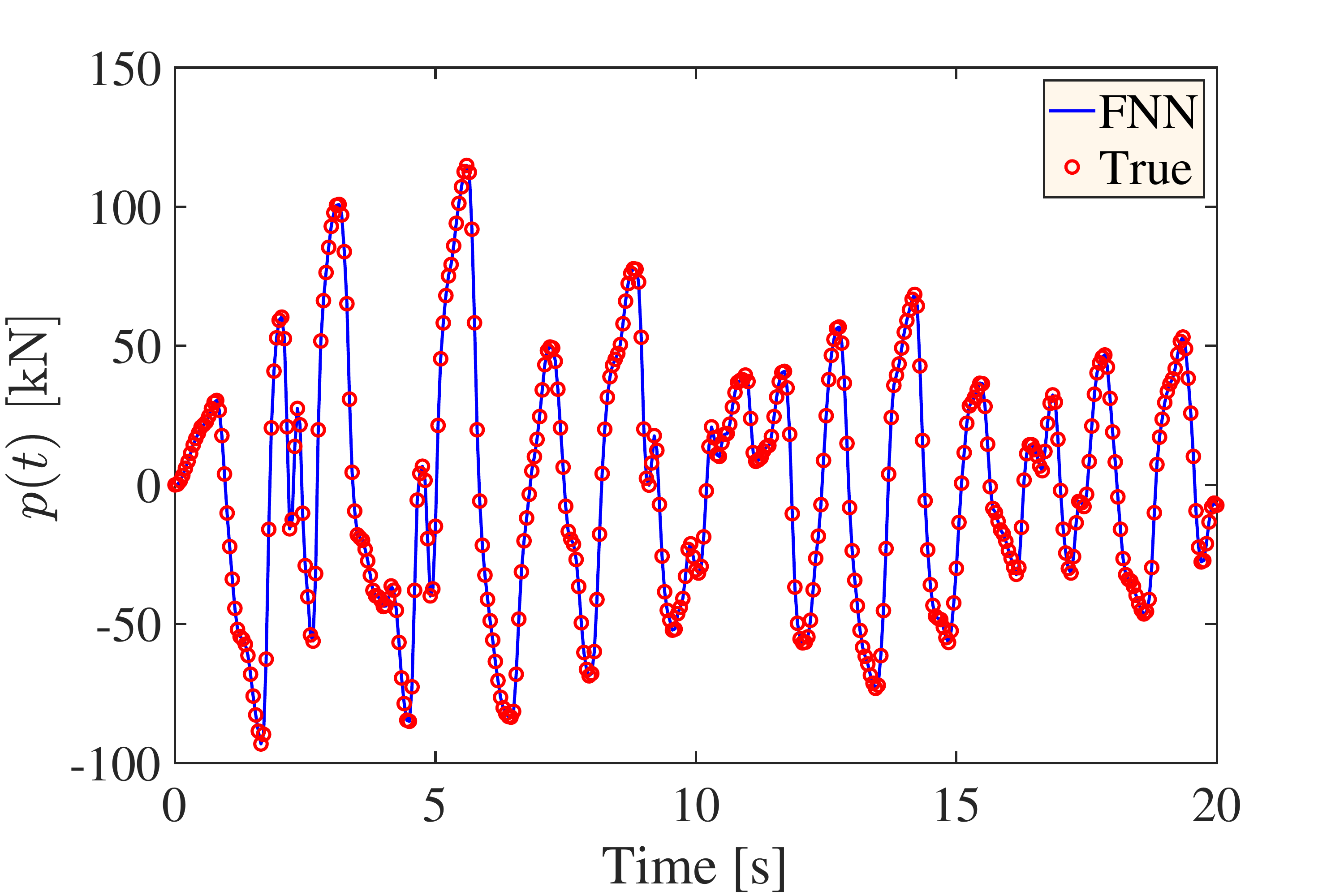}
    \caption{Prediction of $p(t)$ using FNN}
    \end{subfigure}
    \hfill
    \begin{subfigure}[t]{0.48\textwidth}
    \centering
    \includegraphics[scale=0.3]{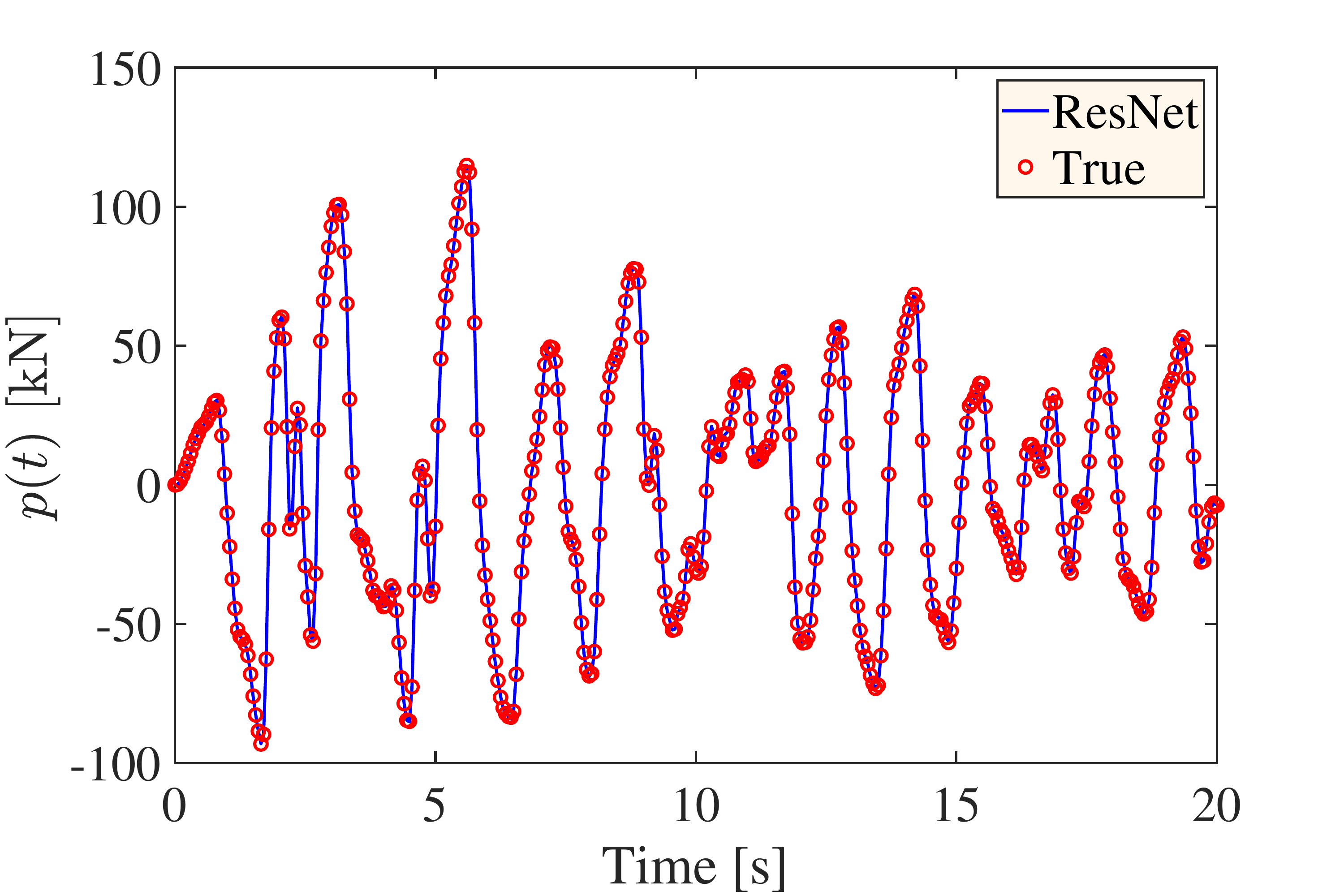}
    \caption{Prediction of $p(t)$ using ResNet}
    \end{subfigure}\\
    \begin{subfigure}[t]{0.48\textwidth}
    \centering
    \includegraphics[scale=0.3]{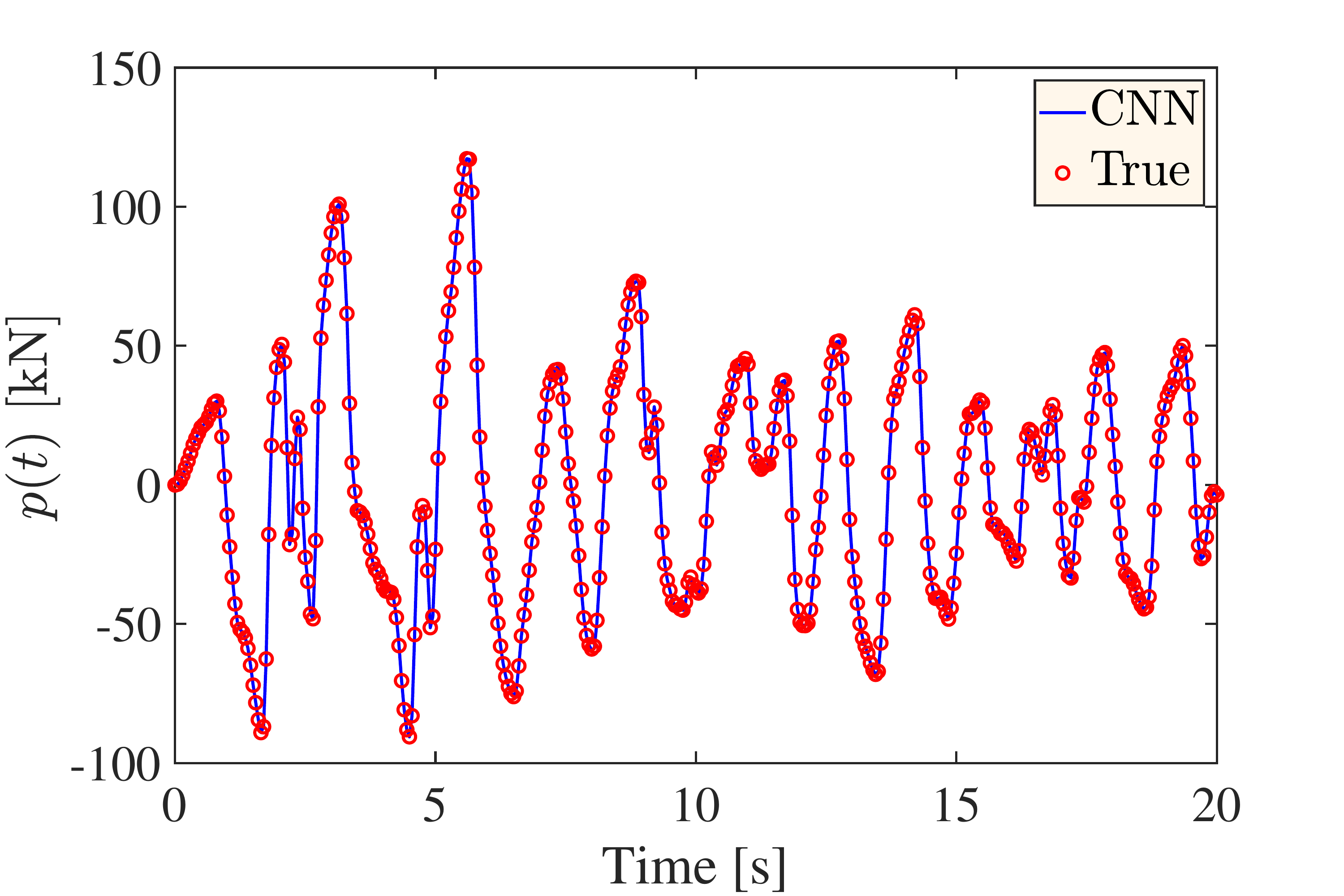}
    \caption{Prediction of $p(t)$ using CNN}
    \end{subfigure}
    \caption{Comparison of predictions from three different neural network architectures for a realization of the uncertainty in the validation dataset $\Dval$ in Example II. The true solution is obtained by solving \eqref{eq:nvie}. }
    \label{fig:100dof_p_valid}
\end{figure}

In this example, the neural networks are used to model the uncertain nonlinear restoring force in the base layer as $p(t;\xii)=f_\mathrm{b}$. The training dataset $\Dtr=\Bigg\{\Big\{p\left(t_i;\xii_j^{(\mathrm{tr})}\right)\Big\}_{i=1}^{\nt},\xii_j^{(\mathrm{tr})}\Bigg\}_{j=1}^{\Ntr}$ is generated using $\Ntr=450$ random samples of the uncertain parameters drawn from their respective probability distributions given in Table \ref{tab:ExII_param} and with a 20 Hz temporal sampling rate. For validation dataset $\Dval=\Bigg\{\Big\{p\left(t_i;\xii_j^{(\mathrm{val})}\right)\Big\}_{i=1}^{\nt},\xii_j^{(\mathrm{val})}\Bigg\}_{j=1}^{\Nval}$, separate $\Nval=50$ random samples are used. A similar procedure to the previous example is followed here to select the number of neurons per layer as $m=100$ and total number of hidden layers as $N_H=4$ in the FNN network. 
The activation function is chosen as the hyperbolic tangent function $\sigma_{\mathrm{tanh}}(\cdot)$ (see \eqref{eq:act1}) as it gives the smallest validation RMSE. ResNet uses a similar configuration with a residual connection between the first and third layer as this residual connection gives the smallest error among other residual connections. 
For CNN, a similar procedure is followed and $N_C=3$ one-dimensional convolution layers with kernels of length three followed by two feed-forward layers with $n_t$ neurons in each of these layers are used. The activation function for the convolution layers are chosen as the hyperbolic tangent function $\sigma_{\mathrm{tanh}}(\cdot)$, whereas the feed-forward layers use the ELU activation $\sigma_{\!_\mathrm{ELU}}(\cdot)$. A learning rate of $10^{-3}$ that is gradually halved every 2000 iterations for training of FNN and ResNet and halved every 500 iterations for training of CNN subjected to a maximum iteration of 10000 is used to train these networks using the Adam algorithm (see Appendix \ref{sec:adam}). 
The training of FNN and ResNet networks takes approximately 3.5 hours, whereas the training of CNN network takes approximately 6 hours. 
The network parameters that produces the smallest validation RMSE is selected as the trained network at the end of training, which is equivalent to an early stopping criterion \cite{bengio2012practical}. 
The validation RMSE using these three architectures are given in Table \ref{tab:100_val_rmse}, which shows that the CNN provides the smallest validation RMSE.  Figure \ref{fig:100dof_p_valid} shows the estimated $p(t;\xii)=f_\mathrm{b}$ using these architectures for one realization of the uncertain parameters in the validation dataset $\Dval$. 

\begin{figure}
    \centering
    \begin{subfigure}[t]{\textwidth}
    \centering
    \includegraphics[scale=0.8]{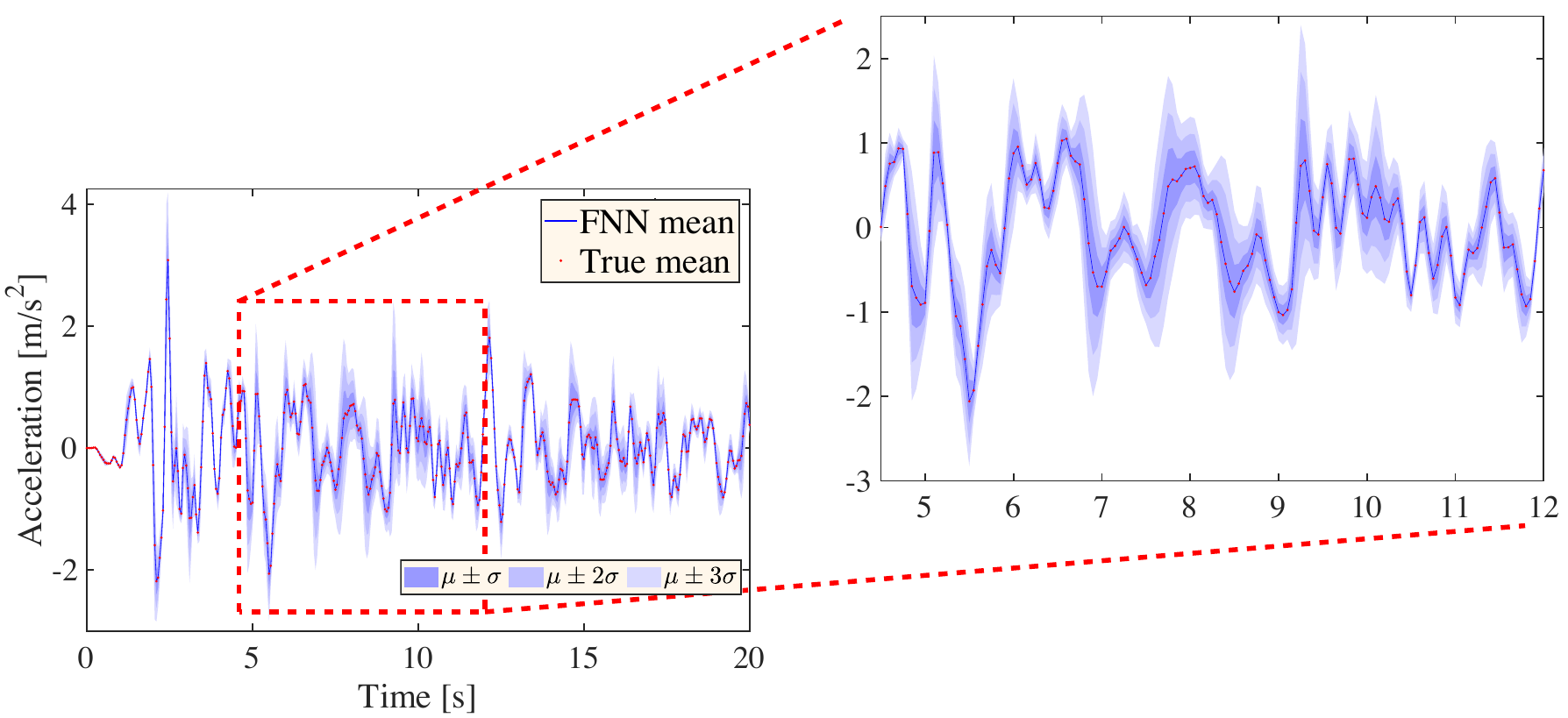}
    \caption{Prediction of mean and standard deviation of the horizontal roof acceleration using FNN}
    \end{subfigure}\\
    \begin{subfigure}[t]{\textwidth}
    \centering
    \includegraphics[scale=0.8]{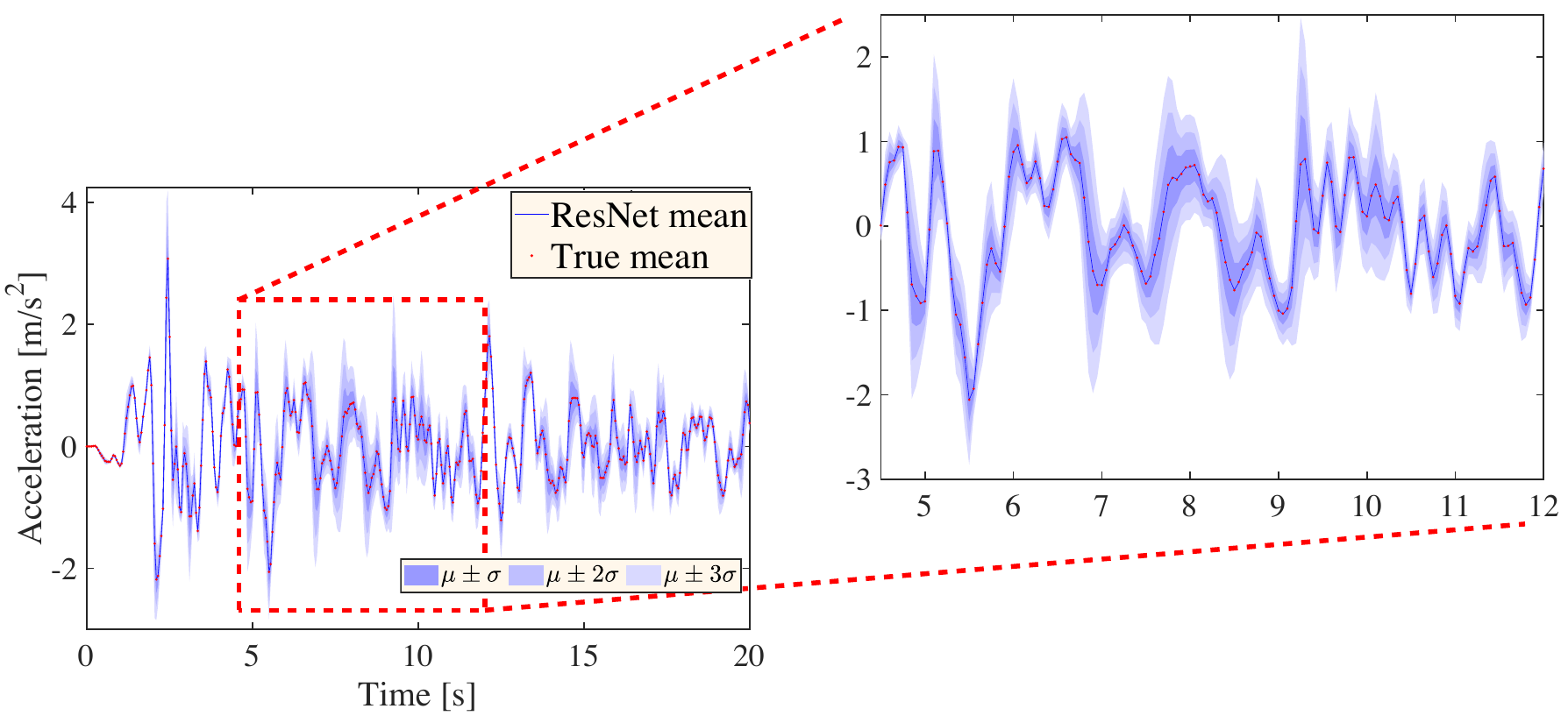}
    \caption{Prediction of mean and standard deviation of the horizontal roof acceleration using ResNet}
    \end{subfigure}
    \begin{subfigure}[t]{\textwidth}
    \centering
    \includegraphics[scale=0.8]{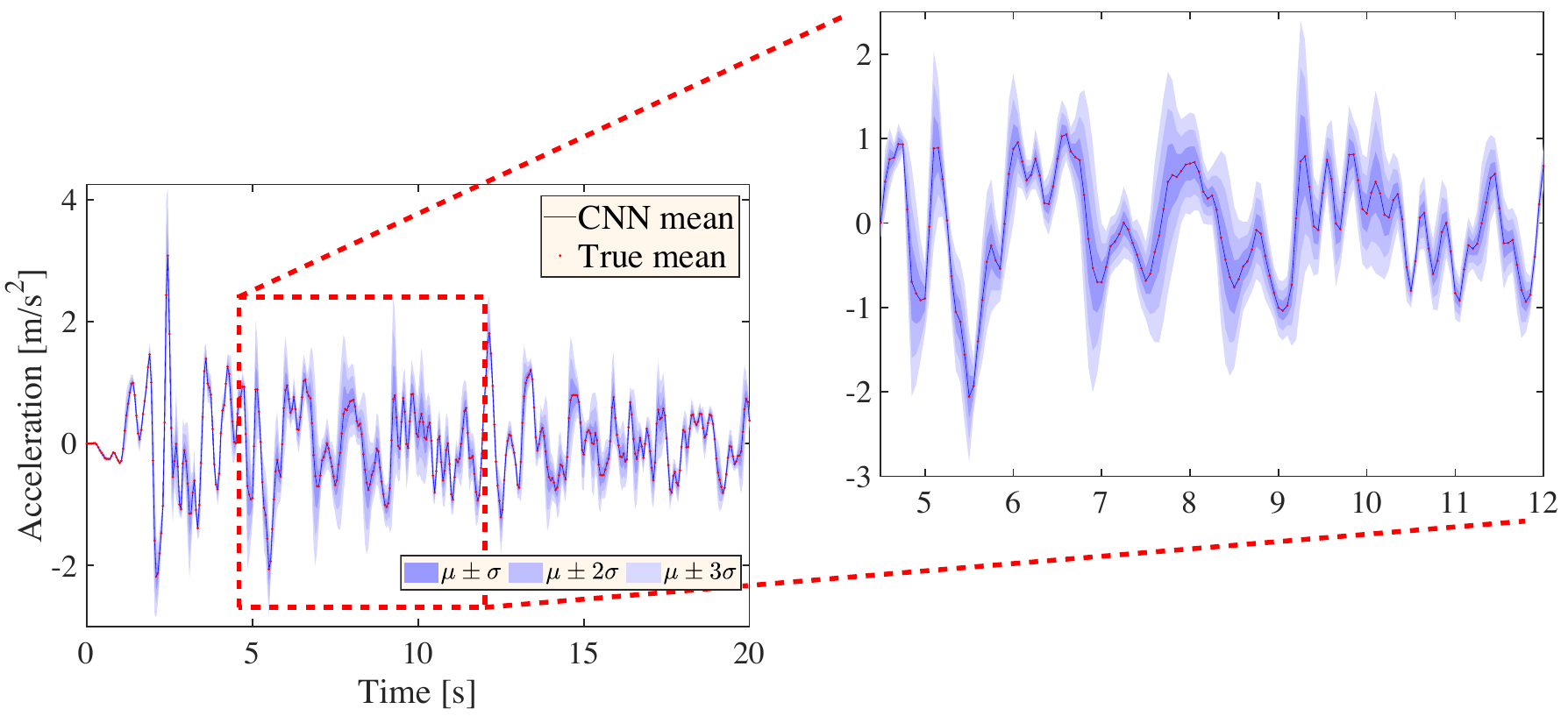}
    \caption{Prediction of mean and standard deviation of the horizontal roof acceleration using CNN}
    \end{subfigure}
    \caption{Prediction of the mean and standard deviations of the horizontal roof acceleration $\ddot{u}_{97}$ from three different neural network architectures using $10^5$ realizations of the uncertainty in Example II with response in $[4.5,12]$ s zoomed in. The mean responses from the neural networks are compared with mean estimated from solving \eqref{eq:nvie} for $10^5$ realizations of the uncertainty. The shaded regions show combinations of mean $\mu$ and standard deviation $\sigma$ of the response.}
    \label{fig:100dof_accel}
\end{figure}

\begin{figure}[!htb]
    \centering
    \includegraphics[scale=1]{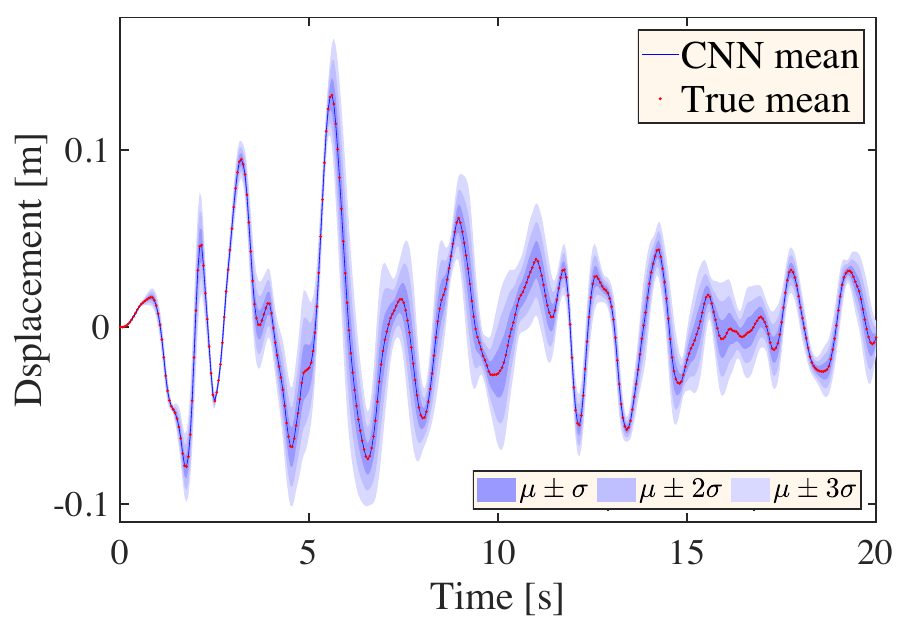}
    \caption{Prediction of the mean and standard deviations of horizontal roof displacement $u_{97}$ from three different neural network architectures using $10^5$ realizations of the uncertainty in Example I. The mean responses from the neural networks are compared with mean estimated from solving \eqref{eq:nvie} for $10^5$ realizations of the uncertainty. The shaded regions show combinations of mean $\mu$ and standard deviation $\sigma$ of the response.} 
    \label{fig:100dof_pred_disp} 
\end{figure}

Figure \ref{fig:100dof_accel} compares the mean of the horizontal roof acceleration $\ddot{u}_{97}$ for $N=10^5$ random samples of the uncertain parameters in Table \ref{tab:ExII_param} obtained from the proposed use of the neural networks with the mean estimated from solving \eqref{eq:nvie}. 
The figure also shows the estimated standard deviation of the response. Table \ref{tab:100_pred_rmse} shows the RMSEs in the mean and standard deviation estimates. Similar to the previous example CNN produces the smallest error as it avoids overfitting by parameter sharing as described in Section \ref{sec:cnn}. On the other hand, ResNet gives validation RMSE one order of magnitude larger than CNN showing that modeling the residual does not provide any advantage in this approach. The same trained CNN is further used to estimate the mean and standard deviation of the roof displacement $u_{97}$ with $N=10^5$ realizations of the uncertain parameters (see Figure \ref{fig:100dof_pred_disp}). 
Once trained the FNN, ResNet, and CNN take a total $1.13$ hrs., $1.13$ hrs., and $1.12$ hrs., respectively, for predicting responses for $10^5$ different realizations of the uncertainty. 
On the other hand, in this example solving \eqref{eq:nvie} using a computationally efficient method \cite{de2018computationally} takes $15.33$ hrs. in total for these $10^5$ evaluations. A standard nonlinear solver, \textsc{Matlab}'s \texttt{ode45}, however, takes $12.98$ s for one evaluation and hence it would take approximately $15$ days for $10^5$ evaluations if this solver is used. 

\begin{table}[!htb]
\caption{Validation RMSE for the isolator force $f_\mathrm{b}$ using three different architectures for the neural networks in Example II. } 
\centering 
\begin{tabular}{c c c c} 
\hline 
\Tstrut
NN architecture & Validation RMSE \\ [0.5ex] 
\hline 
\Tstrut
 FNN & $2.8954\times10^{-3}$ \\ 
 ResNet & $1.2112\times10^{-2}$ \\
 CNN & $1.4874\times10^{-3}$ \\ [1ex] 
\hline 
\end{tabular}
\label{tab:100_val_rmse} 
\end{table}

\begin{table}[!htb]
\caption{RMSE of mean and standard deviation of the horizontal roof acceleration $\ddot{u}_{97}$ using three different architectures for the neural networks in Example II. } 
\centering 
\begin{tabular}{c c c c} 
\hline 
\Tstrut
NN architecture & RMSE of mean & RMSE of std dev \\ [0.5ex] 
\hline 
\Tstrut
 FNN & $1.1369\times10^{-3}$ & $8.5926\times10^{-3}$\\ 
 ResNet & $1.2870\times10^{-3}$ & $8.7527\times10^{-3}$\\
 CNN & $1.0753\times10^{-3}$ & $8.1803\times10^{-3}$\\ [1ex] 
\hline 
\end{tabular}
\label{tab:100_pred_rmse} 
\end{table}

\subsection{Example III: 1623 Degree-of-freedom Three-Dimensional Wind-Excited Structure}

\begin{figure}[!htb]
    \centering
    \includegraphics[scale=0.65]{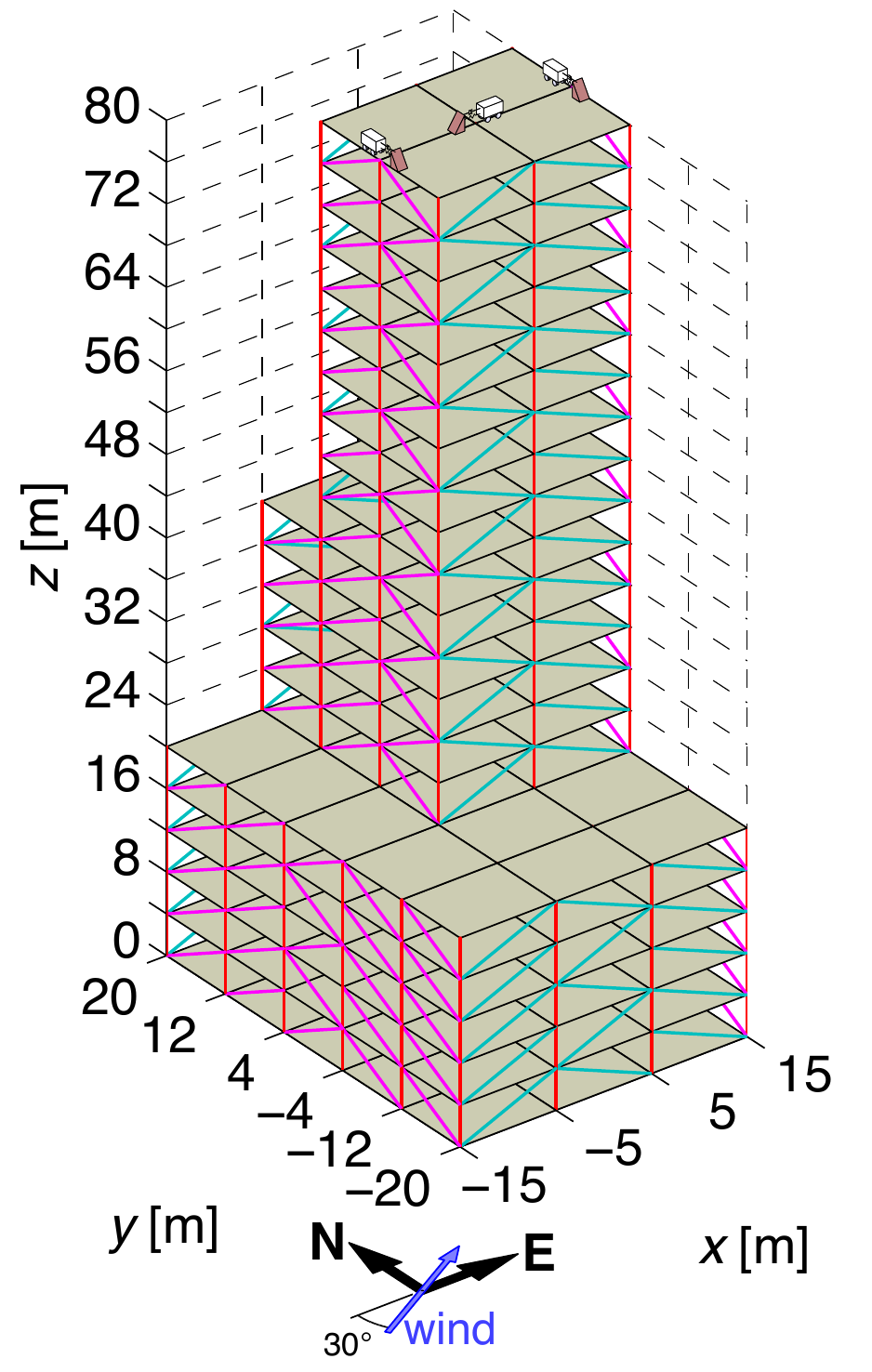}
    \caption{A $20$-story building model with three tuned mass dampers at its roof used in Example III.}
    \label{fig:1623dof}
\end{figure}

The third example uses a $20$-story moment-resisting building frame with $1623$ DOF adapted from \cite{wojtkiewicz2014efficient}, where three nonlinear Tuned Mass Dampers (TMDs) are attached to its roof. The dimensions of the building are shown in Figure \ref{fig:1623dof}. Its bottom five stories have $5\times3$ bay. However, next five stories are reduced to $3\times2$ bay and the last ten stories to $2\times2$ bay. The building has cross braces to provide extra stiffness against lateral bending and torsion. The beams and columns in the building are modeled using the Euler-Bernoulli beam theory and the beam-column joints as rigid. 
The building has 1620 DOF without the TMDs and its first six natural frequencies are summarized in Table \ref{tab:1623_freq}. 
\begin{table}[!htb]
\caption{First six natural frequencies of the $1623$ DOF building used in Example III. } 
\centering 
\begin{tabular}{c c c c} 
\hline 
\Tstrut
Mode No. & Mode type & Frequency (Hz) \\ [0.5ex] 
\hline 
\Tstrut
$1^\mathrm{st}$ & $y$-direction  & $0.5718$ \\ 
$2^\mathrm{nd}$ & $x$-direction  & $0.5893$ \\
$3^\mathrm{rd}$ & torsional  & $0.9363$ \\
$4^\mathrm{th}$ & $y$-direction  & $1.3632$ \\
$5^\mathrm{th}$ & $x$-direction  & $1.5346$ \\ 
$6^\mathrm{th}$ & torsional  & $2.0292$ \\ [1ex] 
\hline 
\end{tabular}
\label{tab:1623_freq} 
\end{table}
The building is subjected to a wind excitation from the northeast direction at an angle of $30^\circ$ from east, which is modeled as a narrowband filtered Gaussian white noise filtered through a $16^\mathrm{th}$ order band-pass Butterworth filter with cutoff frequencies set at 1.2 times smaller and larger than the fundamental natural frequency. This choice of wind load excites the fundamental modes in the $x$- and $y$-directions, and in torsion. 
The wind excitation is shaped according to a power law model that is proportional to its height to the power 0.3 \cite{holmes1996along}. 
The damping forces in the three TMDs are assumed to follow a power law model given by 
\begin{equation}
    g(\dot{u}) = f_\mathrm{TMD} = {c}_1 \dot{u} + {c}_2\lvert \dot{u} \rvert^\beta \mathrm{sgn}(\dot{u}),
\end{equation}
where $u$ is the displacement of the TMD relative to the roof; and the damping coefficients $c_1$, $c_2$, and the exponent $\beta$ are assumed uncertain with their distributions shown in Table \ref{tab:ExIII_param}. 

\begin{center}
\begin{table}[!htb]
\caption{Probability distribution of the uncertain parameters in Example III. } \label{tab:ExIII_param} 
\centering
\begin{tabular}{l c c c c c c} 
\hline 
\Tstrut
\multirow{2}{*}{Parameter} & \multirow{2}{*}{Distribution} & \multicolumn{2}{c}{$x$ TMD} & \multicolumn{2}{c}{$y$ TMDs} \\\cline {3-4} \cline{5-6} \Tstrut 
& & Mean & Std. Dev. & Mean & Std. Dev. \\[0.5ex] 
\hline 
\Tstrut
$c_1$ [kN$\cdot$s/m] & Gaussian & $225$ & $150$ & $120$ & $80$  \\ 
$c_2$ [kN$\cdot$(s/m)$^\beta$] & Lognormal & $27.5$ & $12.5$ & $20.0$ & $10.0$ \\
$\beta$ & Lognormal & $0.85$ & $0.20$ & $0.85$ & $0.20$ \\ [1ex] 
\hline 
\end{tabular}
\end{table}
\end{center}

\subsubsection{Results}

\begin{figure}[!htb]
    \centering
    \begin{subfigure}[t]{0.48\textwidth}
    \centering
    \includegraphics[scale=0.3]{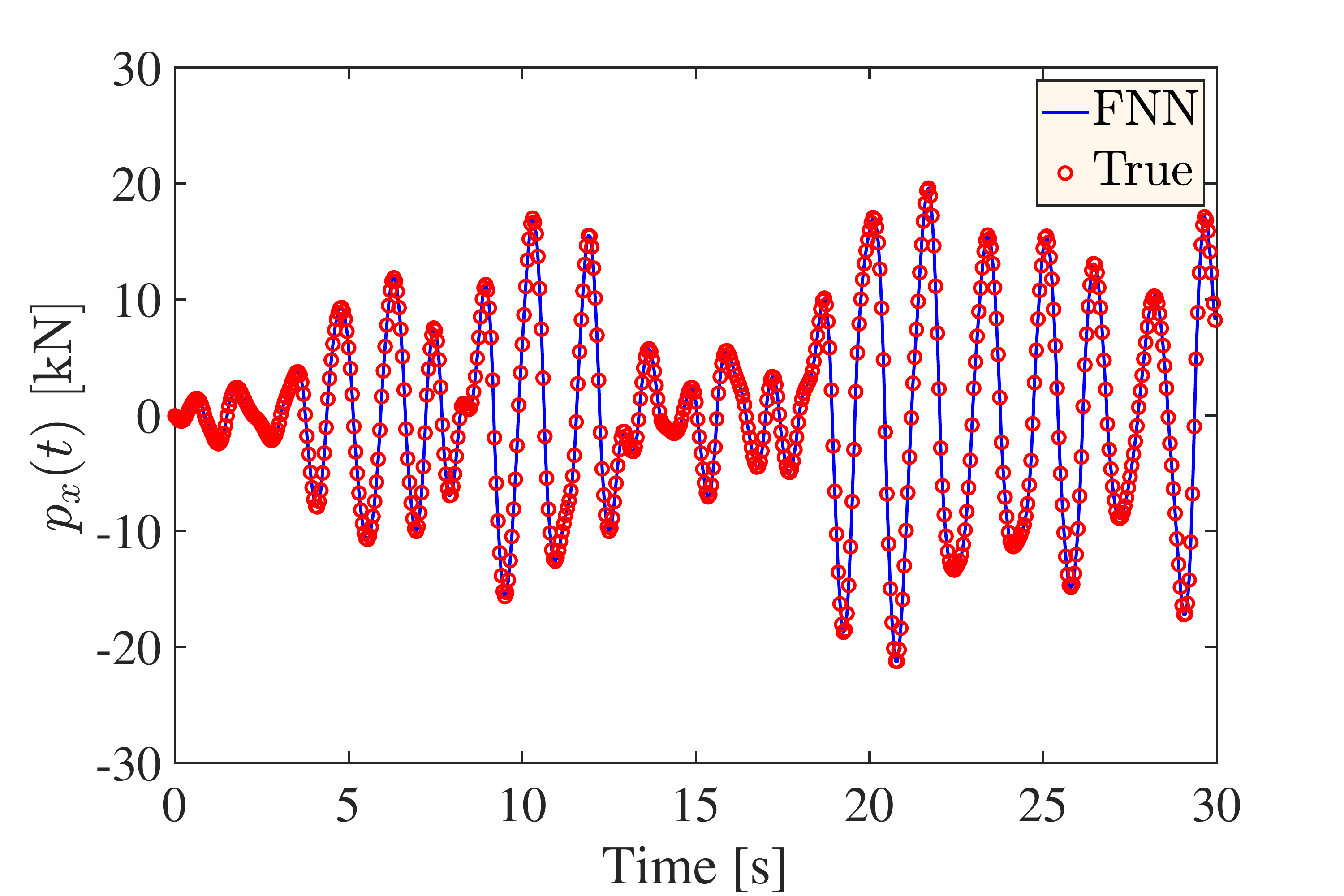}
    \caption{$x$-direction TMD force predicted by FNN}
    \end{subfigure}
    \hfill
    \begin{subfigure}[t]{0.48\textwidth}
    \centering
    \includegraphics[scale=0.3]{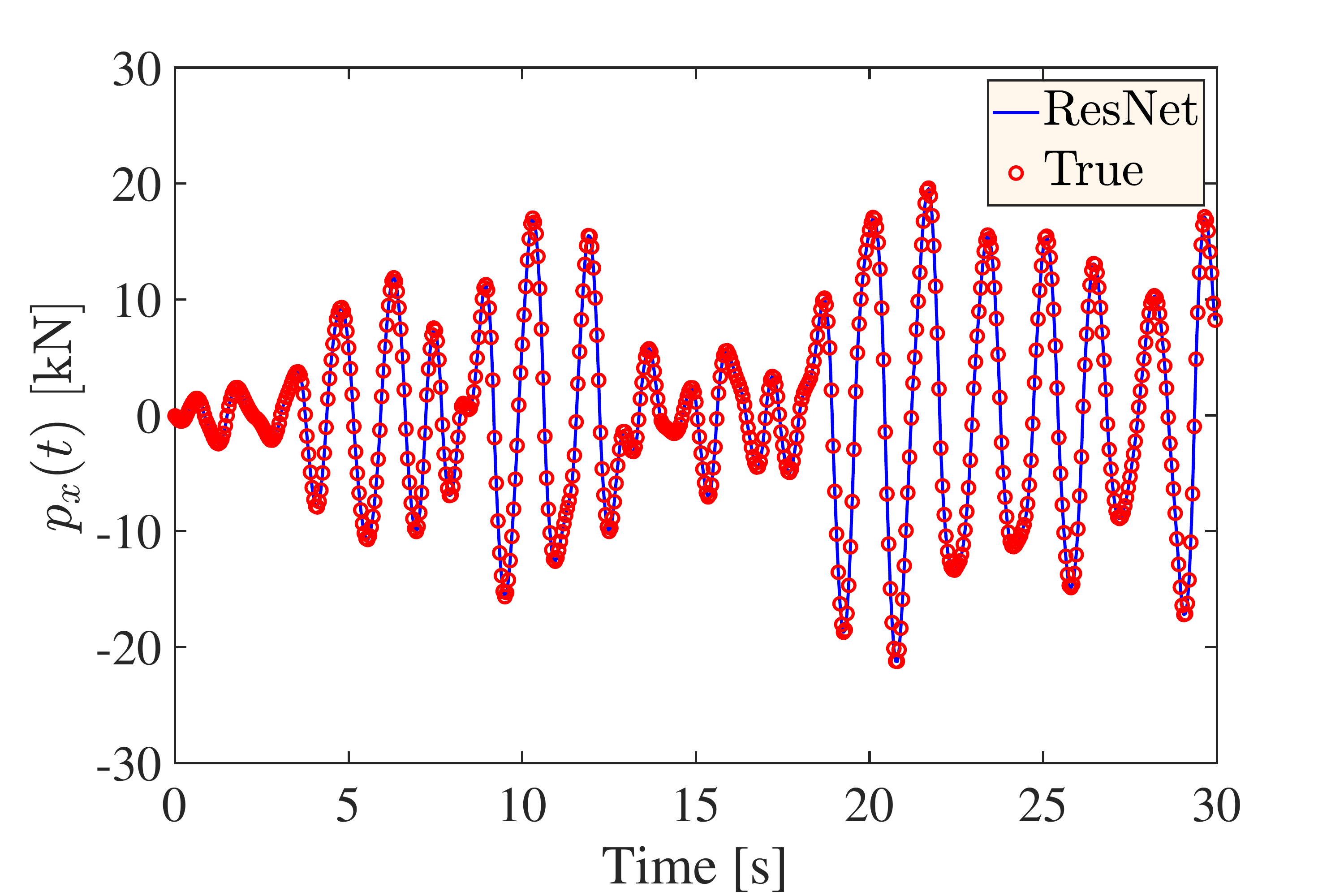}
    \caption{$x$-direction TMD force predicted by ResNet}
    \end{subfigure}\\
    \begin{subfigure}[t]{0.48\textwidth}
    \centering
    \includegraphics[scale=0.3]{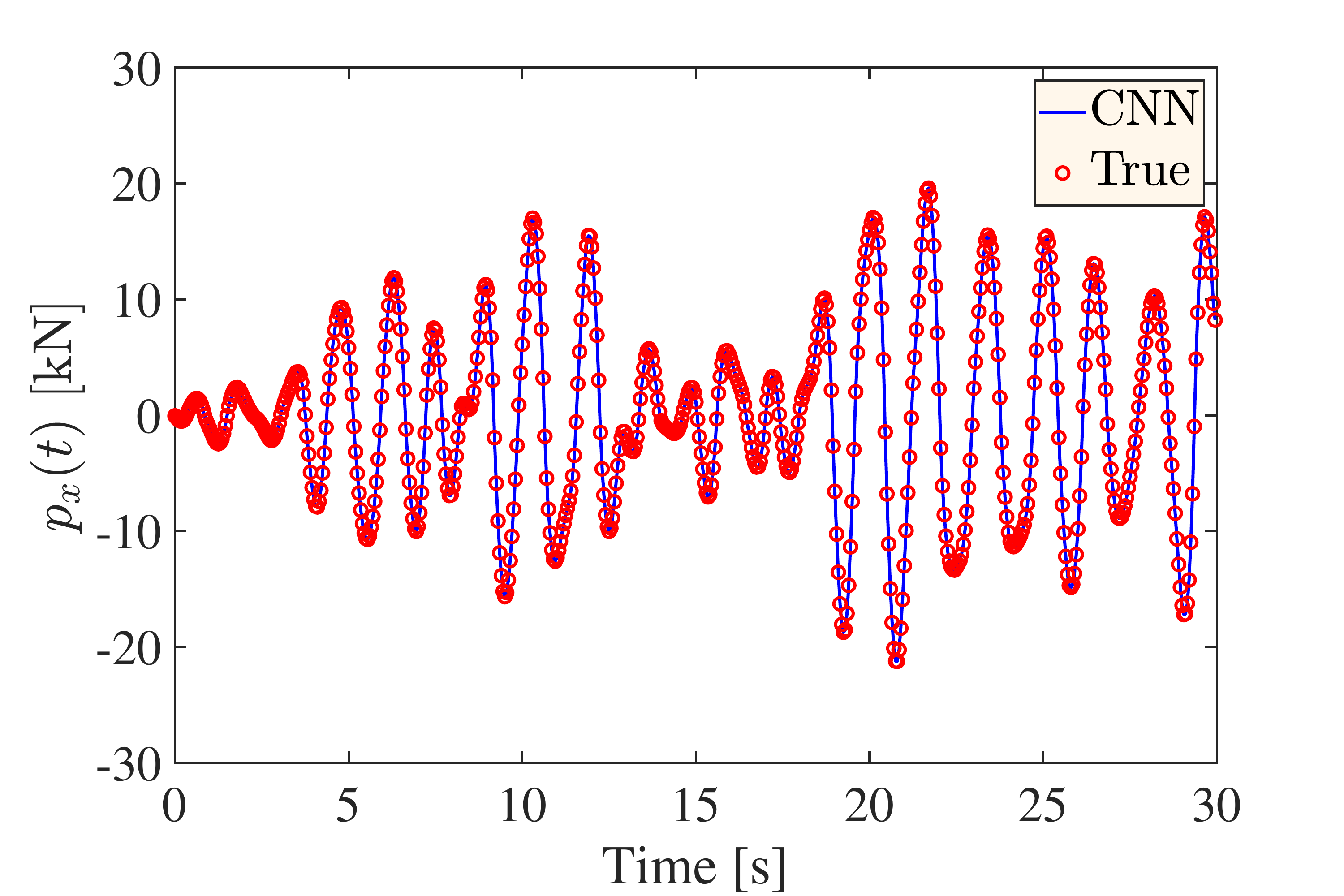}
    \caption{$x$-direction TMD force predicted by CNN}
    \end{subfigure}
    \caption{Comparison of predictions of the $x$-direction TMD force from three different neural network architectures for a realization of the uncertainty in the validation dataset $\Dval$ in Example III. The true solution is obtained by solving \eqref{eq:nvie}.}
    \label{fig:1623_p_valid}
\end{figure}

In this example, the neural networks are used to model the uncertain nonlinear forces in the TMDs on the roof as $p(t;\xii)=f_\mathrm{TMD}$. The training dataset $\Dtr=\Bigg\{\Big\{p\left(t_i;\xii_j^{(\mathrm{tr})}\right)\Big\}_{i=1}^{\nt},\xii_j^{(\mathrm{tr})}\Bigg\}_{j=1}^{\Ntr}$ is generated using $\Ntr=250$ random samples of the uncertain parameters drawn from their respective probability distributions given in Table \ref{tab:ExIII_param} and with a 20 Hz temporal sampling rate. For validation dataset $\Dval=\Bigg\{\Big\{p\left(t_i;\xii_j^{(\mathrm{val})}\right)\Big\}_{i=1}^{\nt},\xii_j^{(\mathrm{val})}\Bigg\}_{j=1}^{\Nval}$, separate $\Nval=50$ random samples are used. A similar procedure to the previous example is followed here to select the number of neurons per layer $m=200$ and total number of hidden layers $N_H=4$ in the FNN network. 
The activation function is chosen as the hyperbolic tangent function $\sigma_{\mathrm{tanh}}(\cdot)$ (see \eqref{eq:act1}) as it gives the smallest validation RMSE. ResNet uses a similar configuration with a residual connection between the first and third layer as other residual connections do not provide smaller validation RMSE. 
For CNN, a similar procedure is followed and $N_C=3$ one-dimensional convolution layers with kernels of length three followed by two feed-forward layers with $n_t$ neurons in each of these layers are used. The activation function for the convolution layers are chosen as the hyperbolic tangent function $\sigma_{\mathrm{tanh}}(\cdot)$, whereas the feed-forward layers use the ELU activation $\sigma_{\!_\mathrm{ELU}}(\cdot)$. These networks are trained using the Adam algorithm (see Appendix \ref{sec:adam}) with a learning rate of $10^{-3}$, which is gradually halved every 2000 iterations for training of FNN and ResNet but halved every 500 iterations for training of CNN subjected to a maximum iteration of 10000. The training of FNN and ResNet networks takes approximately 4 hours each, whereas the training of CNN networks takes approximately 6 hours each. 
The network parameters that produces the smallest validation RMSE is selected as the trained network at the end of training, which is equivalent to an early stopping criterion \cite{bengio2012practical}. 
The validation RMSE using these three architectures are given in Table \ref{tab:1623_val_rmse}, which shows that the CNN provides the smallest validation RMSE as it prevents overfitting by parameter sharing. Figure \ref{fig:1623_p_valid} shows that the estimated force  $p_x(t;\xii)=f^x_\mathrm{TMD}$ in the $x$-direction TMD using these architectures for one realization of the uncertain parameters in the validation dataset $\Dval$ matches the true values. 

\begin{table}[!htb]
\caption{Validation RMSE for the TMD forces using three different architectures for the neural networks in Example III. } 
\centering 
\begin{tabular}{c c c c} 
\hline 
\Tstrut
TMD & NN architecture & Validation RMSE \\ [0.5ex] 
\hline 
\Tstrut
\multirow{3}{*}{$x$-direction TMD} & FNN & $6.3901\times10^{-4}$ \\ 
 & ResNet & $3.0472\times10^{-3}$ \\
 & CNN & $2.0155\times10^{-4}$ \\[0.5ex] \hline \Tstrut
\multirow{3}{*}{$y$-direction TMD\#1} & FNN & $8.2597\times10^{-4}$ \\ 
 & ResNet & $2.9137\times10^{-3}$ \\
 & CNN & $4.3792\times10^{-4}$ \\[0.5ex] \hline \Tstrut
 \multirow{3}{*}{$y$-direction TMD\#2} & FNN & $1.2655\times10^{-3}$ \\ 
 & ResNet & $1.8165\times10^{-3}$ \\
 & CNN & $6.4929\times10^{-4}$ \\ [1ex] 
\hline 
\end{tabular}
\label{tab:1623_val_rmse} 
\end{table}

\begin{table}[!htb]
\caption{RMSE of mean and standard deviation of the roof accelerations in $x$- and $y$-directions using three different architectures for the neural networks in Example III.} 
\centering 
\begin{tabular}{c c c c} 
\hline 
\Tstrut
Roof acceleration & NN architecture & RMSE of mean & RMSE of std dev \\ [0.5ex] 
\hline 
\Tstrut
\multirow{3}{*}{$x$-direction} & FNN & $2.6617\times10^{-4}$ & $4.9477\times10^{-3}$ \\ 
 & ResNet & $7.4273\times10^{-4}$ & $2.1737\times10^{-2}$ \\
 & CNN & $2.6451\times10^{-4}$ & $3.6935\times10^{-3}$ \\[0.5ex] \hline \Tstrut
\multirow{3}{*}{$y$-direction} & FNN & $2.2522\times10^{-4}$ & $1.2896\times10^{-2}$ \\ 
 & ResNet & $3.1833\times10^{-3}$ & $2.0854\times10^{-2}$ \\
 & CNN & $5.3049\times10^{-4}$ & $1.1705\times10^{-2}$ \\ [1ex] 
\hline 
\end{tabular}
\label{tab:1623_pred_rmse} 
\end{table}

\begin{figure}
    \centering
    \begin{subfigure}[t]{\textwidth}
    \centering
    \includegraphics[scale=0.8]{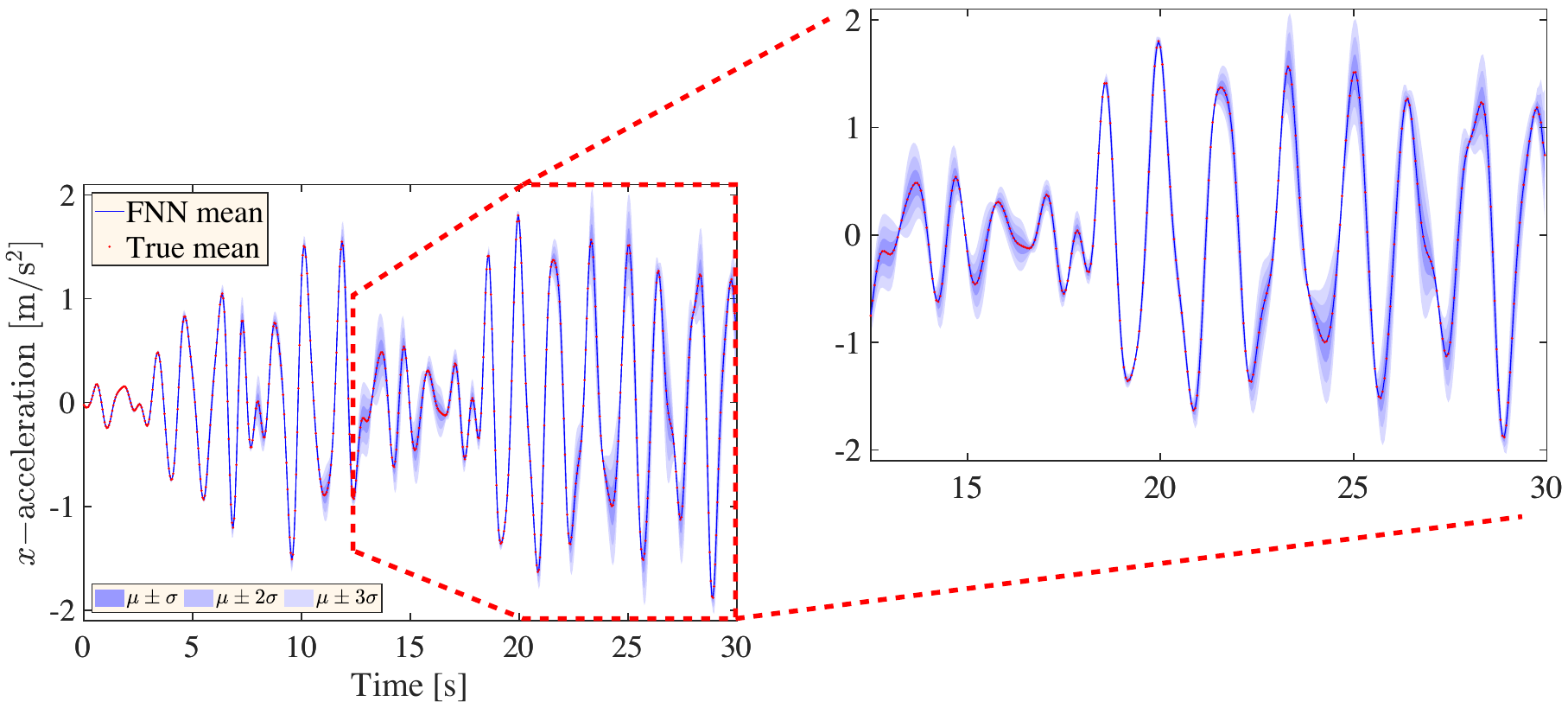}
    \caption{Prediction of mean and standard deviation of the roof acceleration in $x$-direction using FNN}
    \end{subfigure}\\
    \begin{subfigure}[t]{\textwidth}
    \centering
    \includegraphics[scale=0.8]{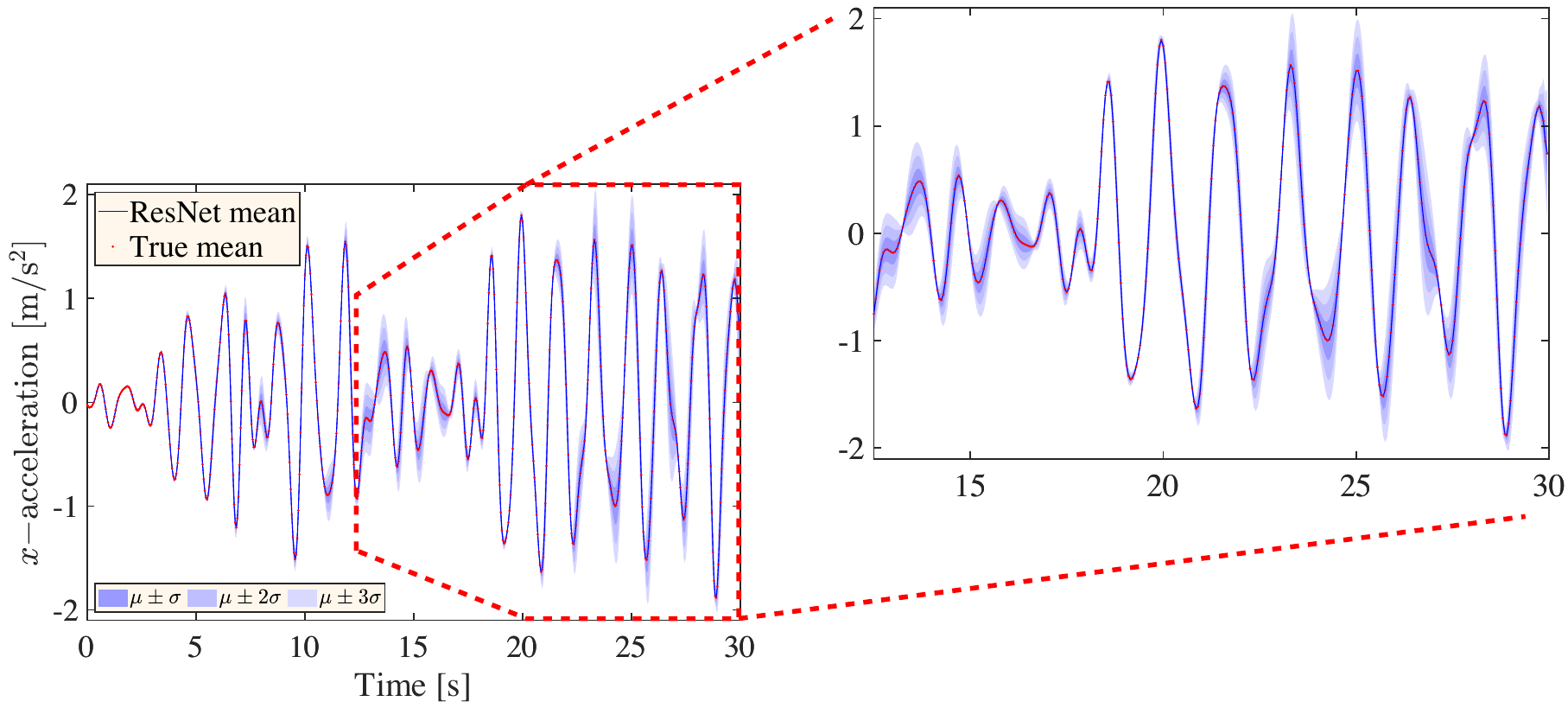}
    \caption{Prediction of mean and standard deviation of the roof acceleration in $x$-direction using ResNet}
    \end{subfigure}
    \begin{subfigure}[t]{\textwidth}
    \centering
    \includegraphics[scale=0.8]{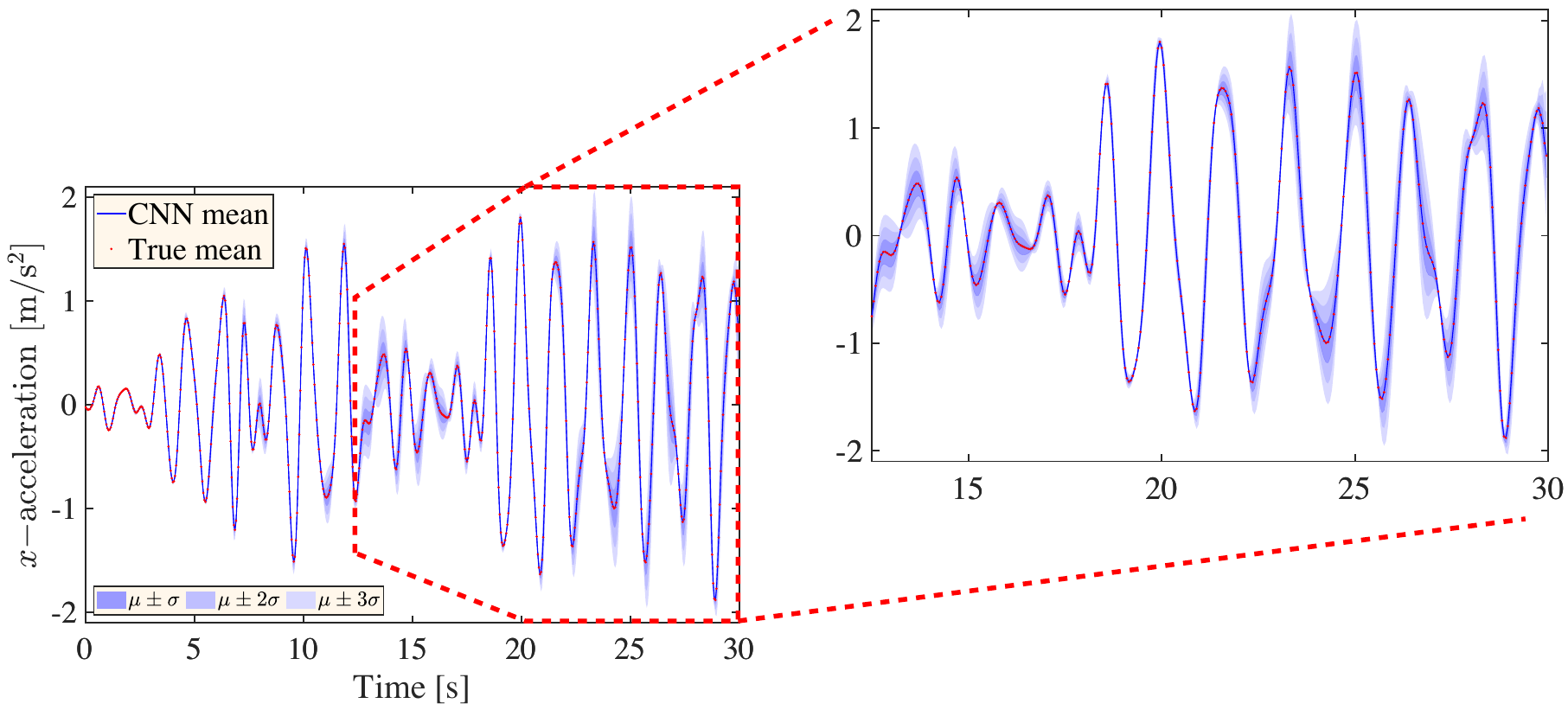}
    \caption{Prediction of mean and standard deviation of the roof acceleration in $x$-direction using CNN}
    \end{subfigure}
    \caption{Prediction of the mean and standard deviations of the $x$-direction acceleration at the center of the roof from three different neural network architectures using $10^5$ realizations of the uncertainty in Example III with response in $[12.5,30]$ s zoomed in. The mean responses from the neural networks are compared with mean estimated from solving \eqref{eq:nvie} for $10^5$ realizations of the uncertainty. The shaded regions show combinations of mean $\mu$ and standard deviation $\sigma$ of the response.}
    \label{fig:1623dof_x_accel}
\end{figure}

\begin{figure}
    \centering
    \begin{subfigure}[t]{\textwidth}
    \centering
    \includegraphics[scale=0.8]{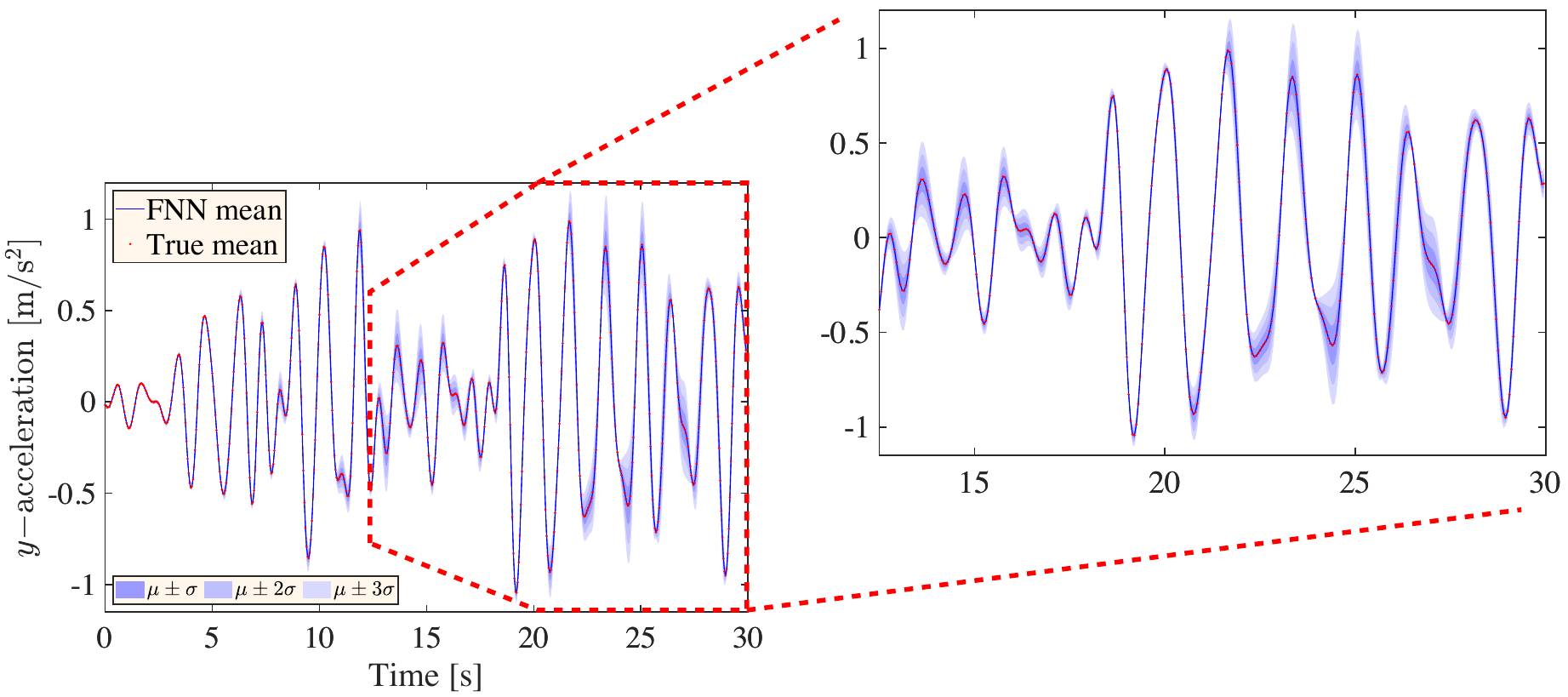}
    \caption{Prediction of mean and standard deviation of the roof acceleration in $y$-direction using FNN}
    \end{subfigure}\\
    \begin{subfigure}[t]{\textwidth}
    \centering
    \includegraphics[scale=0.8]{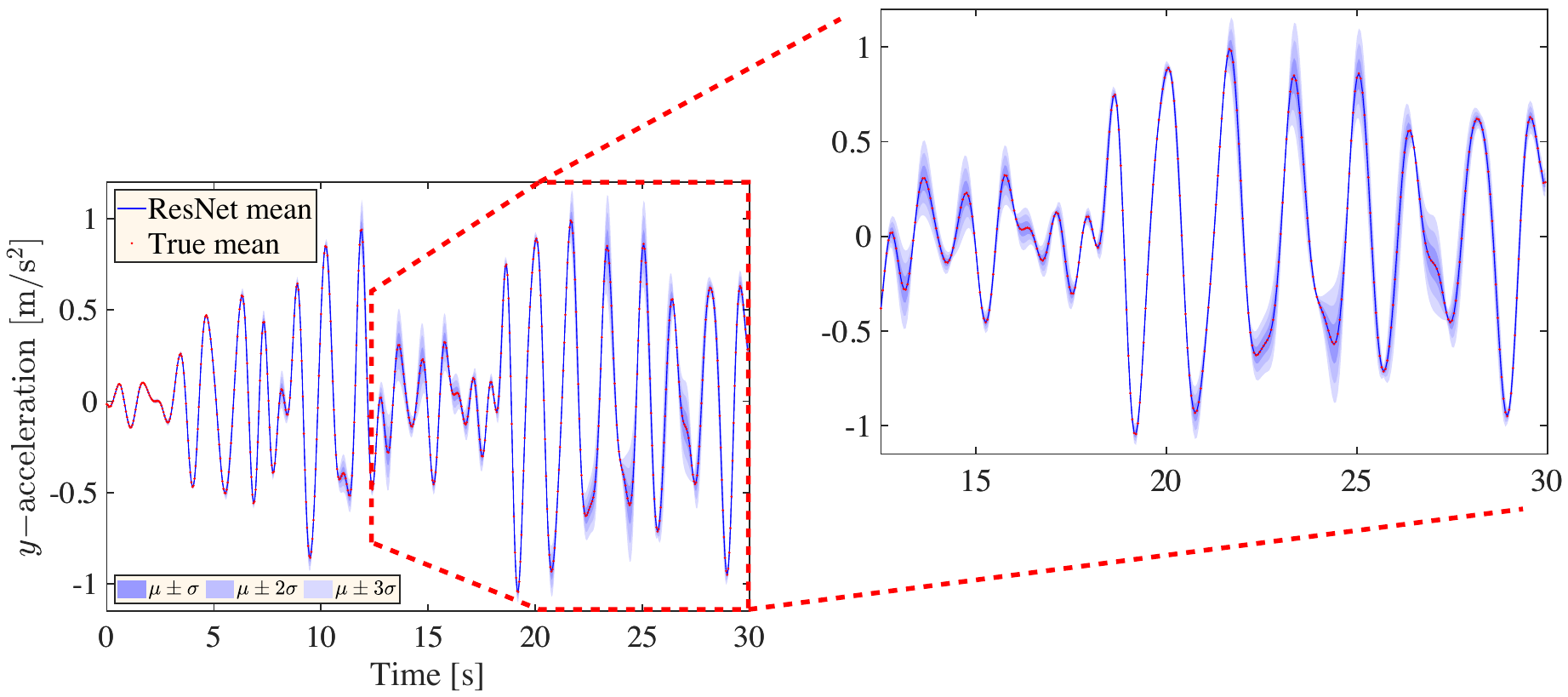}
    \caption{Prediction of mean and standard deviation of the roof acceleration in $y$-direction using ResNet}
    \end{subfigure}
    \begin{subfigure}[t]{\textwidth}
    \centering
    \includegraphics[scale=0.8]{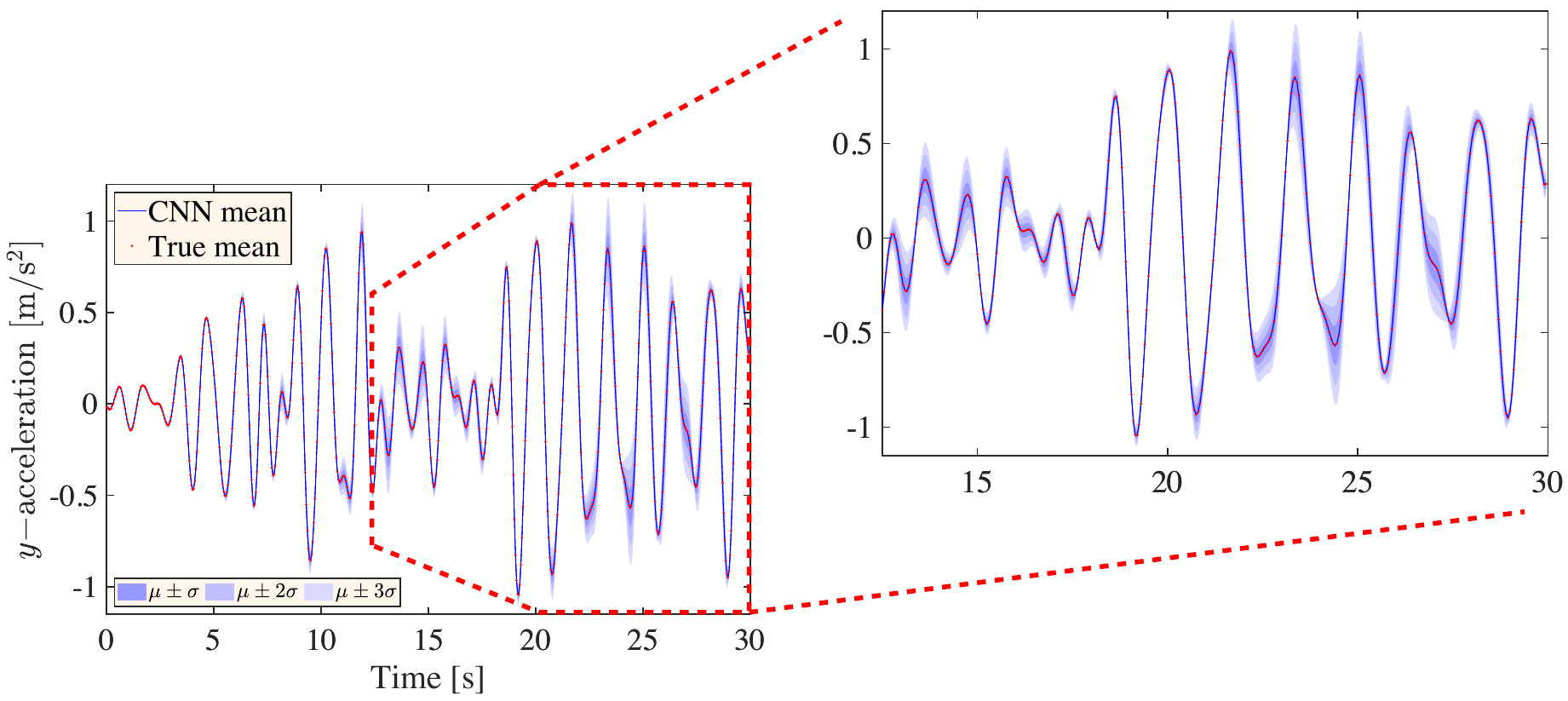}
    \caption{Prediction of mean and standard deviation of the roof acceleration in $y$-direction using CNN}
    \end{subfigure}
    \caption{Prediction of the mean and standard deviations of the $y$-direction acceleration at the center of the roof from three different neural network architectures using $10^5$ realizations of the uncertainty in Example III with response in $[12.5,30]$ s zoomed in. The mean responses from the neural networks are compared with mean estimated from solving \eqref{eq:nvie} for $10^5$ realizations of the uncertainty. The shaded regions show combinations of mean $\mu$ and standard deviation $\sigma$ of the response.}
    \label{fig:1623dof_y_accel}
\end{figure}


Figure \ref{fig:1623dof_x_accel} and \ref{fig:1623dof_y_accel} compare the mean of the roof acceleration in the $x$- and $y$-directions, respectively, for $N=10^5$ random samples of the uncertain parameters in Table \ref{tab:ExIII_param} obtained from the proposed use of neural networks with mean estimated from solving \eqref{eq:nvie}. 
The figures also show the estimated standard deviation of the responses. Table \ref{tab:1623_pred_rmse} shows the RMSEs in the mean and standard deviation estimates of the roof accelerations in both $x$- and $y$-directions. 
Similar to the previous example, CNN produces the smallest error as it avoids overfitting by parameter sharing as described in Section \ref{sec:cnn}. On the other hand, ResNet gives validation RMSE much larger than CNN showing that modeling the residual does not provide any advantage in this method. The RMSE in standard deviation is also higher as it is a difficult statistic to estimate compared to the mean. Note that the same trained networks are used to estimate the mean and standard deviation of accelerations in both $x$- and $y$-directions. 
Once trained the FNN, ResNet, and CNN take a total $2.96$ hrs., $2.97$ hrs., and $2.92$ hrs., respectively, for predicting responses from $10^5$ different realizations of the uncertainty. 
On the other hand, in this example solving \eqref{eq:nvie} using a computationally efficient method \cite{de2018computationally} takes $30.23$ hrs. for these evaluations. A standard nonlinear solver, \textsc{Matlab}'s \texttt{ode45}, however, takes $14.20$ min. for one evaluation and hence it would take approximately $2.74$ yrs. (projected) for $10^5$ evaluations if this solver is used, which is impractical. Note that the same desktop as in the previous examples is used to compute the computational timings. 

\section{Conclusions}
In structural engineering, local nonlinearities often exist in structures like buildings, bridges, or spacecrafts. For these dynamical systems, the computational cost of uncertainty quantification can be significantly large if a nonlinear solver is used. In this paper, the response of a locally nonlinear dynamcial system is divided into response of a nominal linear system and response from a pseudoforce that arises from the uncertainty and local nonlinearity. 
Recently, neural networks have become popular for representing a functional relationship due to their high level of expressiveness. Further, with the availability of advanced computational resources and open-sourced packages like PyTorch and TensorFlow the training of neural networks have become possible on a desktop. In this paper, three different architectures of these neural networks are investigated to predict the pseudoforce in the second part of the response from the nonlinearity and uncertainty present in the system. Three numerical examples with DOF ranging from two to 1623 are used to illustrate the efficacy of the proposed approach. These examples show that the neural networks can accurately model the pseudoforces and the total response of the system. Once trained these neural networks are efficiently used for estimating statistics of the response under uncertainty. The computational efficiency will be further pronounced for more complex engineering systems, which will be investigated in future. 

\appendix
\section{Adam Algorithm \cite{kingma2014adam}} \label{sec:adam}
In the Adam algorithm, historical gradient information is used to retard the descent along large gradients \cite{kingma2014adam,de2019topology}. This information is stored in $\widehat \mm$ and $\widehat \vm$ as
\begin{equation}
\begin{split}
\mm_{k} &= b_m \mm_{k-1} + (1-b_m) \frac{\partial J}{\partial \thetaa_{k}}; \quad\widehat{\mm}_{k} = \frac{\mm^{(k)}}{1-b_m^k};\\
\vm_{k} &= b_v \vm_{k-1}+(1-b_v)\left[\frac{\partial J}{\partial \thetaa_{k}}\right]^2; \quad \widehat{\vm}_{k}=\frac{\vm^{(k)}}{1-b_v^k},\\
\end{split}
\end{equation}
where $\left[\frac{\partial J}{\partial \thetaa_{k}}\right]^2$ is performed element-wise; and $b_m$ and $b_v$ are parameters of the Adam algorithm with default values 0.9 and 0.999, respectively. The gradient descent step is applied next as follows
\begin{equation}
\thetaa_{k+1} = \thetaa_{k} - \eta \frac{\widehat{\mm}_{k}}{\sqrt{\widehat{\vm}_{k}}+\epsilon},
\end{equation}
where the above update is performed element-wise and $\epsilon$ is a very small number to avoid division by zero. We use this algorithm to train the neural networks in this paper. An illustration of the steps of this algorithm is shown in Algorithm \ref{alg:adam}.

\begin{algorithm}
	\begin{algorithmic}
	    \State \textbf{Network}: $\pinn(\cdot;\thetaa_0)$
		\State \textbf{Given}: $\eta$, $b_m$, $b_v$, and $\epsilon$.
		\Procedure{Adam}{}
		\State Initialize $\thetaa_{1}=\thetaa_0$.
        \State Initialize $\mm_0 = \mathbf{0}$.
        \State Initialize $\vm_0 = \mathbf{0}$.
		\For {$k=1,2,\dots,N_{\max}$}
		\State Compute $\frac{\partial J}{\partial \thetaa_{k}}$.
		\State Set $\mm_k \leftarrow b_m\mm_{k-1} + (1-b_m)\frac{\partial J}{\partial \thetaa_{k}}$. 
		\State Set $\vm_k \leftarrow b_v\vm_{k-1} + (1-b_v)\left[\frac{\partial J}{\partial \thetaa_{k}}\right]^2\qquad$(element-wise). 
		\State Set $\widehat\mm_k \leftarrow \mm_k/(1-b_m^k)$. 
		\State Set $\widehat\vm_k \leftarrow \vm_k/(1-b_v^k)$. 
		\State Set $\thetaa_{k+1} \leftarrow \thetaa_{k} - \eta\frac{\widehat{\mm}_k}{\sqrt{\widehat{\vm}_k}+{\epsilon}}\qquad $(element-wise). 
		\EndFor
		\EndProcedure
        \State \textbf{Trained network}: $\pinn(\cdot;\thetaa_{\!_{N_{\max}}})$.
	\end{algorithmic}
	\caption{\textit{Adam} \cite{kingma2014adam}}
	\label{alg:adam}
\end{algorithm}

\bibliographystyle{unsrt}  
\bibliography{references}  

\begin{thebibliography}{10}

\bibitem{hasselman2001quantification}
Timothy Hasselman.
\newblock Quantification of uncertainty in structural dynamic models.
\newblock {\em Journal of Aerospace Engineering}, 14(4):158--165, 2001.

\bibitem{uncertainty2006}
Kurt Maute and Chris~L Pettit.
\newblock Uncertainty quantification and design under uncertainty of aerospace
  systems.
\newblock {\em Structure and Infrastructure Engineering}, 2(3-4):159--159,
  2006.

\bibitem{bulleit2008uncertainty}
William~M Bulleit.
\newblock Uncertainty in structural engineering.
\newblock {\em Practice Periodical on Structural Design and Construction},
  13(1):24--30, 2008.

\bibitem{ghanem2003stochastic}
Roger~G Ghanem and Pol~D Spanos.
\newblock {\em Stochastic finite elements: a spectral approach}.
\newblock Courier Corporation, 2003.

\bibitem{xiu2002wiener}
Dongbin Xiu and George~Em Karniadakis.
\newblock The {W}iener--{A}skey polynomial chaos for stochastic differential
  equations.
\newblock {\em SIAM journal on scientific computing}, 24(2):619--644, 2002.

\bibitem{babuvska2007stochastic}
Ivo Babu{\v{s}}ka, Fabio Nobile, and Raul Tempone.
\newblock A stochastic collocation method for elliptic partial differential
  equations with random input data.
\newblock {\em SIAM Journal on Numerical Analysis}, 45(3):1005--1034, 2007.

\bibitem{nobile2008sparse}
Fabio Nobile, Ra{\'u}l Tempone, and Clayton~G Webster.
\newblock A sparse grid stochastic collocation method for partial differential
  equations with random input data.
\newblock {\em SIAM Journal on Numerical Analysis}, 46(5):2309--2345, 2008.

\bibitem{williams2006gaussian}
Christopher~KI Williams and Carl~Edward Rasmussen.
\newblock {\em Gaussian processes for machine learning}.
\newblock MIT press Cambridge, MA, 2006.

\bibitem{forrester2007multi}
Alexander~IJ Forrester, Andr{\'a}s S{\'o}bester, and Andy~J Keane.
\newblock Multi-fidelity optimization via surrogate modelling.
\newblock {\em Proceedings of the royal society a: mathematical, physical and
  engineering sciences}, 463(2088):3251--3269, 2007.

\bibitem{romero2004construction}
VJ~Romero, LP~Swiler, and AA~Giunta.
\newblock Construction of response surfaces based on
  progressive-lattice-sampling experimental designs with application to
  uncertainty propagation.
\newblock {\em Structural Safety}, 26(2):201--219, 2004.

\bibitem{giunta2006promise}
AA~Giunta, JM~McFarland, LP~Swiler, and MS~Eldred.
\newblock The promise and peril of uncertainty quantification using response
  surface approximations.
\newblock {\em Structures and Infrastructure Engineering}, 2(3-4):175--189,
  2006.

\bibitem{mckay2000comparison}
Michael~D McKay, Richard~J Beckman, and William~J Conover.
\newblock A comparison of three methods for selecting values of input variables
  in the analysis of output from a computer code.
\newblock {\em Technometrics}, 42(1):55--61, 2000.

\bibitem{ross1990course}
Sheldon~M Ross.
\newblock {\em Simulation}.
\newblock Academic Press, 5th edition, 2013.

\bibitem{kamalzare2015efficient}
Mahmoud Kamalzare, Erik~A Johnson, and Steven~F Wojtkiewicz.
\newblock Efficient optimal design of passive structural control applied to
  isolator design.
\newblock {\em Smart Structures and Systems}, 15(3):847--862, 2015.

\bibitem{de2015fast}
Subhayan De, Erik~A Johnson, and Steven~F Wojtkiewicz.
\newblock Fast {B}ayesian model selection with application to large
  locally-nonlinear dynamic systems.
\newblock In {\em 6th International Conference on Advances in Experimental
  Structural Engineering, 11th International Workshop on Advanced Smart
  Materials and Smart Structures Technology}, 2015.

\bibitem{de2018computationally}
Subhayan De, Erik~A Johnson, Steven~F Wojtkiewicz, and Patrick~T Brewick.
\newblock Computationally efficient {B}ayesian model selection for locally
  nonlinear structural dynamic systems.
\newblock {\em Journal of Engineering Mechanics}, 144(5):04018022, 2018.

\bibitem{de2017efficient}
Subhayan De, Steven~F Wojtkiewicz, and Erik~A Johnson.
\newblock Efficient optimal design and design-under-uncertainty of passive
  control devices with application to a cable-stayed bridge.
\newblock {\em Structural Control and Health Monitoring}, 24(2):e1846, 2017.

\bibitem{ferri1988modeling}
Aldo~A Ferri.
\newblock Modeling and analysis of nonlinear sleeve joints of large space
  structures.
\newblock {\em Journal of Spacecraft and Rockets}, 25(5):354--360, 1988.

\bibitem{bowden1990joint}
Mary Bowden and John Dugundji.
\newblock Joint damping and nonlinearity in dynamics of space structures.
\newblock {\em AIAA Journal}, 28(4):740--749, 1990.

\bibitem{krack2017vibration}
Malte Krack, Loic Salles, and Fabrice Thouverez.
\newblock Vibration prediction of bladed disks coupled by friction joints.
\newblock {\em Archives of Computational Methods in Engineering},
  24(3):589--636, 2017.

\bibitem{ying1992analysis}
Ren Ying.
\newblock {\em The analysis and identification of friction joint parameters in
  the dynamic response of structures}.
\newblock PhD thesis, 1992.

\bibitem{lee2004dynamic}
Yongsik Lee and ZC~Feng.
\newblock Dynamic responses to sinusoidal excitations of beams with frictional
  joints.
\newblock {\em Communications in Nonlinear Science and Numerical Simulation},
  9(6):571--581, 2004.

\bibitem{poudou2007modeling}
Olivier Poudou.
\newblock {\em Modeling and analysis of the dynamics of dry-friction-damped
  structural systems}.
\newblock PhD thesis, 2007.

\bibitem{wojtkiewicz2011efficient}
Gaurav, SF~Wojtkiewicz, and EA~Johnson.
\newblock Efficient uncertainty quantification of dynamical systems with local
  nonlinearities and uncertainties.
\newblock {\em Probabilistic Engineering Mechanics}, 26(4):561--569, 2011.

\bibitem{de2015efficienta}
Subhayan De, Steven~F Wojtkiewicz, and Erik~A Johnson.
\newblock Efficient optimal design-under-uncertainty of passive structural
  control devices.
\newblock In {\em Proceedings of the 12th International Conference on
  Applications of Statistics and Probability in Civil Engineering (ICASP12)},
  2015.

\bibitem{de2015efficientb}
Subhayan De, Erik~A Johnson, Steven~F Wojtkiewicz, and Patrick~T Brewick.
\newblock Efficient {B}ayesian model selection for identifying locally
  nonlinear systems incorporating dynamic measurements.
\newblock In F.-K. Chang and F.~Kopsaftopoulos, editors, {\em Structural Health
  Monitoring 2015: System reliability for verification and implementation},
  volume~2, page 2318–2325. DEStech Publications, Lancaster, PA, 2015.

\bibitem{paszke2017automatic}
Paszke Adam, Gross Sam, Chintala Soumith, Chanan Gregory, Yang Edward,
  D~Zachary, Lin Zeming, Desmaison Alban, Antiga Luca, and Lerer Adam.
\newblock Automatic differentiation in {PyTorch}.
\newblock In {\em Proceedings of Neural Information Processing Systems}, 2017.

\bibitem{abadi2016tensorflow}
Mart{\'\i}n Abadi, Paul Barham, Jianmin Chen, Zhifeng Chen, Andy Davis, Jeffrey
  Dean, Matthieu Devin, Sanjay Ghemawat, Geoffrey Irving, Michael Isard, et~al.
\newblock Tensorflow: A system for large-scale machine learning.
\newblock In {\em 12th {USENIX} symposium on operating systems design and
  implementation ({OSDI} 16)}, pages 265--283, 2016.

\bibitem{baker2019workshop}
Nathan Baker, Frank Alexander, Timo Bremer, Aric Hagberg, Yannis Kevrekidis,
  Habib Najm, Manish Parashar, Abani Patra, James Sethian, Stefan Wild, Karen
  Willcox, and Steven Lee.
\newblock Workshop report on basic research needs for scientific machine
  learning: Core technologies for artificial intelligence.
\newblock Technical report, USDOE Office of Science (SC), Washington, DC
  (United States), 2019.

\bibitem{raissi2017physics}
Maziar Raissi, Paris Perdikaris, and George~Em Karniadakis.
\newblock Physics informed deep learning (part {I}): {Data-driven} solutions of
  nonlinear partial differential equations.
\newblock {\em arXiv preprint arXiv:1711.10561}, 2017.

\bibitem{raissi2018hidden}
Maziar Raissi and George~Em Karniadakis.
\newblock Hidden physics models: Machine learning of nonlinear partial
  differential equations.
\newblock {\em Journal of Computational Physics}, 357:125--141, 2018.

\bibitem{raissi2018deep}
Maziar Raissi.
\newblock Deep hidden physics models: Deep learning of nonlinear partial
  differential equations.
\newblock {\em The Journal of Machine Learning Research}, 19(1):932--955, 2018.

\bibitem{raissi2019physics}
Maziar Raissi, Paris Perdikaris, and George~E Karniadakis.
\newblock Physics-informed neural networks: A deep learning framework for
  solving forward and inverse problems involving nonlinear partial differential
  equations.
\newblock {\em Journal of Computational Physics}, 378:686--707, 2019.

\bibitem{lu2019deepxde}
Lu~Lu, Xuhui Meng, Zhiping Mao, and George~E Karniadakis.
\newblock {DeepXDE: A} deep learning library for solving differential
  equations.
\newblock {\em arXiv preprint arXiv:1907.04502}, 2019.

\bibitem{king2018deep}
Ryan King, Oliver Hennigh, Arvind Mohan, and Michael Chertkov.
\newblock From deep to physics-informed learning of turbulence: Diagnostics.
\newblock {\em arXiv preprint arXiv:1810.07785}, 2018.

\bibitem{goodfellow2016nips}
Ian Goodfellow.
\newblock {NIPS} 2016 tutorial: Generative adversarial networks.
\newblock {\em arXiv preprint arXiv:1701.00160}, 2016.

\bibitem{stengel2019physics}
Karen Stengel, Andrew Glaws, and Ryan King.
\newblock Physics-informed super resolution of climatological wind and solar
  resource data.
\newblock {\em AGUFM}, 2019:A43E--04, 2019.

\bibitem{hesthaven2018non}
Jan~S Hesthaven and Stefano Ubbiali.
\newblock Non-intrusive reduced order modeling of nonlinear problems using
  neural networks.
\newblock {\em Journal of Computational Physics}, 363:55--78, 2018.

\bibitem{wang2019non}
Qian Wang, Jan~S Hesthaven, and Deep Ray.
\newblock Non-intrusive reduced order modeling of unsteady flows using
  artificial neural networks with application to a combustion problem.
\newblock {\em Journal of Computational Physics}, 384:289--307, 2019.

\bibitem{holland2019towards}
Jonathan~R Holland, James~D Baeder, and Karthik Duraisamy.
\newblock Towards integrated field inversion and machine learning with embedded
  neural networks for {RANS} modeling.
\newblock In {\em AIAA Scitech 2019 Forum}, page 1884, 2019.

\bibitem{bhatnagar2019prediction}
Saakaar Bhatnagar, Yaser Afshar, Shaowu Pan, Karthik Duraisamy, and Shailendra
  Kaushik.
\newblock Prediction of aerodynamic flow fields using convolutional neural
  networks.
\newblock {\em Computational Mechanics}, 64(2):525--545, 2019.

\bibitem{wang2019prediction}
Jian-Xun Wang, Junji Huang, Lian Duan, and Heng Xiao.
\newblock Prediction of reynolds stresses in high-mach-number turbulent
  boundary layers using physics-informed machine learning.
\newblock {\em Theoretical and Computational Fluid Dynamics}, 33(1):1--19,
  2019.

\bibitem{zhang2019physics}
Ruiyang Zhang, Yang Liu, and Hao Sun.
\newblock Physics-guided convolutional neural network {(PhyCNN)} for
  data-driven seismic response modeling.
\newblock {\em arXiv preprint arXiv:1909.08118}, 2019.

\bibitem{zhang2019deep}
Ruiyang Zhang, Zhao Chen, Su~Chen, Jingwei Zheng, Oral
  B{\"u}y{\"u}k{\"o}zt{\"u}rk, and Hao Sun.
\newblock Deep long short-term memory networks for nonlinear structural seismic
  response prediction.
\newblock {\em Computers \& Structures}, 220:55--68, 2019.

\bibitem{de2020transfer}
Subhayan De, Jolene Britton, Matthew Reynolds, Ryan Skinner, Kenneth Jansen,
  and Alireza Doostan.
\newblock On transfer learning of neural networks using bi-fidelity data for
  uncertainty propagation.
\newblock {\em arXiv preprint arXiv:2002.04495}, 2020.

\bibitem{meng2020composite}
Xuhui Meng and George~Em Karniadakis.
\newblock A composite neural network that learns from multi-fidelity data:
  {Application} to function approximation and inverse {PDE} problems.
\newblock {\em Journal of Computational Physics}, 401:109020, 2020.

\bibitem{motamed2019multi}
Mohammad Motamed.
\newblock A multi-fidelity neural network surrogate sampling method for
  uncertainty quantification.
\newblock {\em arXiv preprint arXiv:1909.01859}, 2019.

\bibitem{chakraborty2020transfer}
Souvik Chakraborty.
\newblock Transfer learning based multi-fidelity physics informed deep neural
  network.
\newblock {\em arXiv preprint arXiv:2005.10614}, 2020.

\bibitem{zhang2019quantifying}
Dongkun Zhang, Lu~Lu, Ling Guo, and George~Em Karniadakis.
\newblock Quantifying total uncertainty in physics-informed neural networks for
  solving forward and inverse stochastic problems.
\newblock {\em Journal of Computational Physics}, 397:108850, 2019.

\bibitem{gal2016dropout}
Yarin Gal and Zoubin Ghahramani.
\newblock Dropout as a {Bayesian} approximation: {Representing} model
  uncertainty in deep learning.
\newblock In {\em international conference on machine learning}, pages
  1050--1059, 2016.

\bibitem{hinton2012improving}
Geoffrey~E Hinton, Nitish Srivastava, Alex Krizhevsky, Ilya Sutskever, and
  Ruslan~R Salakhutdinov.
\newblock Improving neural networks by preventing co-adaptation of feature
  detectors.
\newblock {\em arXiv preprint arXiv:1207.0580}, 2012.

\bibitem{luo2019deep}
Xihaier Luo and Ahsan Kareem.
\newblock Deep convolutional neural networks for uncertainty propagation in
  random fields.
\newblock {\em Computer-Aided Civil and Infrastructure Engineering},
  34(12):1043--1054, 2019.

\bibitem{duraisamy2019turbulence}
Karthik Duraisamy, Gianluca Iaccarino, and Heng Xiao.
\newblock Turbulence modeling in the age of data.
\newblock {\em Annual Review of Fluid Mechanics}, 51:357--377, 2019.

\bibitem{chakraborty2020simulation}
Souvik Chakraborty.
\newblock Simulation free reliability analysis: A physics-informed deep
  learning based approach.
\newblock {\em arXiv preprint arXiv:2005.01302}, 2020.

\bibitem{hammersley2013monte}
John Hammersley.
\newblock {\em Monte {C}arlo methods}.
\newblock Springer Science \& Business Media, 2013.

\bibitem{sudret2017surrogate}
Bruno Sudret, Stefano Marelli, and Joe Wiart.
\newblock Surrogate models for uncertainty quantification: An overview.
\newblock In {\em 2017 11th European Conference on Antennas and Propagation
  (EUCAP)}, pages 793--797. IEEE, 2017.

\bibitem{goodfellow2016deep}
Ian Goodfellow, Yoshua Bengio, and Aaron Courville.
\newblock {\em Deep learning}.
\newblock MIT press, 2016.

\bibitem{he2016deep}
Kaiming He, Xiangyu Zhang, Shaoqing Ren, and Jian Sun.
\newblock Deep residual learning for image recognition.
\newblock In {\em Proceedings of the IEEE Conference on Computer Vision and
  Pattern Recognition}, pages 770--778, 2016.

\bibitem{higham2018deep}
Catherine~F Higham and Desmond~J Higham.
\newblock Deep learning: An introduction for applied mathematicians.
\newblock {\em arXiv preprint arXiv:1801.05894}, 2018.

\bibitem{kingma2014adam}
Diederik~P Kingma and Jimmy Ba.
\newblock Adam: A method for stochastic optimization.
\newblock {\em arXiv preprint arXiv:1412.6980}, 2014.

\bibitem{de2019topology}
Subhayan De, Jerrad Hampton, Kurt Maute, and Alireza Doostan.
\newblock Topology optimization under uncertainty using a stochastic
  gradient-based approach.
\newblock {\em arXiv preprint arXiv:1902.04562}, 2019.

\bibitem{de2019bi}
Subhayan De, Kurt Maute, and Alireza Doostan.
\newblock Bi-fidelity stochastic gradient descent for structural optimization
  under uncertainty.
\newblock {\em arXiv preprint arXiv:1911.10420}, 2019.

\bibitem{cybenko1989approximation}
George Cybenko.
\newblock Approximation by superpositions of a sigmoidal function.
\newblock {\em Mathematics of Control, Signals and Systems}, 2(4):303--314,
  1989.

\bibitem{hornik1989multilayer}
Kurt Hornik, Maxwell Stinchcombe, Halbert White, et~al.
\newblock Multilayer feedforward networks are universal approximators.
\newblock {\em Neural Networks}, 2(5):359--366, 1989.

\bibitem{hornik1990universal}
Kurt Hornik, Maxwell Stinchcombe, and Halbert White.
\newblock Universal approximation of an unknown mapping and its derivatives
  using multilayer feedforward networks.
\newblock {\em Neural Networks}, 3(5):551--560, 1990.

\bibitem{ramallo2002smart}
JC~Ramallo, EA~Johnson, and BF~Spencer~Jr.
\newblock 'smart' base isolation systems.
\newblock {\em Journal of Engineering Mechanics}, 128(10):1088--1099, 2002.

\bibitem{lin1987evolutionary}
YK~Lin and Yan Yong.
\newblock Evolutionary {Kanai-Tajimi} earthquake models.
\newblock {\em Journal of Engineering Mechanics}, 113(8):1119--1137, 1987.

\bibitem{bengio2012practical}
Yoshua Bengio.
\newblock Practical recommendations for gradient-based training of deep
  architectures.
\newblock In {\em Neural networks: Tricks of the trade}, pages 437--478.
  Springer, 2012.

\bibitem{wen1976method}
Yi-Kwei Wen.
\newblock Method for random vibration of hysteretic systems.
\newblock {\em Journal of the Engineering Mechanics Division}, 102(2):249--263,
  1976.

\bibitem{ma2004parameter}
F~Ma, H~Zhang, A~Bockstedte, Greg~C Foliente, and P~Paevere.
\newblock Parameter analysis of the differential model of hysteresis.
\newblock {\em J. Appl. Mech.}, 71(3):342--349, 2004.

\bibitem{wojtkiewicz2014efficient}
Steven~F Wojtkiewicz and Erik~A Johnson.
\newblock Efficient sensitivity analysis of structures with local
  modifications. {I}: time domain responses.
\newblock {\em Journal of Engineering Mechanics}, 140(9):04014067, 2014.

\bibitem{holmes1996along}
JD~Holmes.
\newblock Along wind response of lattice towers--{III. Effective} load
  distributions.
\newblock {\em Engineering Structures}, 18(7):489--494, 1996.

\end{thebibliography}






\end{document}